\documentclass[twocolumn,twocolappendix]{aastex63}

\usepackage{amsmath}
\usepackage{txfonts}
\usepackage{xcolor}
\usepackage{graphicx}
\usepackage{gensymb}
\usepackage{booktabs}
\usepackage{bm}
\usepackage{lipsum}
\AtBeginDocument{\mathcode`v=\varv}

\providecommand{\sorthelp}[1]{} 

\newcommand{\HI}{\ifmmode \mathrm{\ion{H}{1}} \else \ion{H}{1} \fi}
\newcommand{\nh}{\ifmmode N_{{\mathrm{H}} \, \mathrm{I}} \else $N_{{\mathrm{H}} \, \mathrm{I}}$\fi} 
\newcommand{\ROHSA}{{\tt ROHSA}}

\newcommand\ab{\bm{a}}

\newcommand\rb{\bm{r}}

\newcommand\mub{\bm{\mu}}
\newcommand\sigmab{\bm{\sigma}}
\newcommand\thetab{\bm{\theta}}

\newcommand{\Trb}{T_{\mathrm b}}
\newcommand{\Trc}{T_{\mathrm c}}
\newcommand{\Trn}{T_{\mathrm n}}
\newcommand{\Trs}{T_{\mathrm s}}
\newcommand{\Tsys}{T_{\mathrm sys}}

\def\GHz{\ifmmode $\,GHz$\else \,GHz\fi}
\def\MJysr{\ifmmode \,$MJy\,sr\mo$\else \,MJy\,sr\mo\fi}
\def\microns{\ifmmode \,\mu$m$\else \,$\mu$m\fi}

\def\kms{\ifmmode $\,km\,s$^{-1}\else \,km\,s$^{-1}$\fi}

\defcitealias{wolfire_neutral_2003}{W03}
\defcitealias{wolfire_neutral_1995}{W95a}
\defcitealias{wolfire_multiphase_1995}{W95b}
\defcitealias{marchal_2019}{M19}

\received{March 7, 2023}
\revised{3 May, 2023}
\accepted{6 May, 2023}
\submitjournal{ApJ}
\shorttitle{Mapping the multiphase structure of \HI\ in LLIV1}
\shortauthors{Vujeva, Marchal, Martin, and Taank}

\graphicspath{{./}{figures/}}
\graphicspath{{./}{figures/}}

\begin{document}
\title{Mapping the multiphase structure of \HI\ in the Low-Latitude Intermediate-Velocity Arch 1}

\correspondingauthor{Luka Vujeva}
\email{luka.vujeva@mail.utoronto.ca}

\author[0000-0001-7697-8361]{Luka Vujeva}
\affiliation{David A. Dunlap Department of Astronomy and Astrophysics, University of Toronto, 60 St. George Street, Toronto, ON M5S 3H8, Canada}
\affiliation{Canadian Institute for Theoretical Astrophysics, University of Toronto, 60 St. George Street, Toronto, ON M5S 3H8, Canada}
\affiliation{Cosmic DAWN Centre (DAWN), Niels Bohr Institute, University of Copenhagen, Jagtvej 128, DK-2200 Copenhagen N}
\affiliation{Niels Bohr Institute, University of Copenhagen, Lyngbyvej 2, DK-2100 Copenhagen Ø}

\author[0000-0002-5501-232X]{Antoine Marchal}
\affiliation{Canadian Institute for Theoretical Astrophysics, University of Toronto, 60 St. George Street, Toronto, ON M5S 3H8, Canada}
\affiliation{Research School of Astronomy \& Astrophysics, Australian National University, Canberra ACT 2610 Australia}

\author[0000-0002-5236-3896]{Peter G. Martin}
\affiliation{Canadian Institute for Theoretical Astrophysics, University of Toronto, 60 St. George Street, Toronto, ON M5S 3H8, Canada}

\author[0000-0001-8461-5552]{Mukesh Taank}
\affiliation{Canadian Institute for Theoretical Astrophysics, University of Toronto, 60 St. George Street, Toronto, ON M5S 3H8, Canada}
\affiliation{Department of Mathematics and Statistics, University of Guelph, 50 Stone Road E., Guelph, ON N1G 2W1, Canada}

\begin{abstract}
    We have analyzed the thermal and turbulent properties of the Low-Latitude Intermediate-Velocity Arch 1 (LLIV1).
    %
    This was accomplished using archival \HI\ emission and absorption data from two 21\,cm line surveys:  GHIGLS  at 9\farcm 4 resolution and DHIGLS at  $1^\prime$ resolution.
    The spectral decomposition code \ROHSA\ was used to model the column density of different thermal phases and also to analyze an absorption measurement
    against the radio source 4C~+66.09. From the latter we found spin temperature $\Trs \sim 75$\,K, cold gas mass fraction $f\sim0.5$, and turbulent sonic Mach number $M_t\sim3.4$. 
    Similar to the absorption line modeling against 4C~+66.09, our best emission line decomposition model has no unstable gas across the whole field of view, suggesting that the thermal condensation and phase transition are not on-going but rather have reached an equilibrium state.
    The cold phase of LLIV1 appears as a collection of elongated filaments that forms a closed structure within the field decomposed. 
    These substructures follow the orientation of the overall large scale cloud, along the diagonal of the GHIGLS field from north-west to south-east (in Galactic coordinates).
    The angular power spectrum of the cold phase is slightly shallower than that of the warm phase, quantifying that the cold phases have relatively more structure on small scales.
    Our spatially resolved map of the cold gas mass fraction in LLIV1 from DHIGLS reveals significant variations 
    spanning the possible range of $f$, with mean and standard deviation 0.33 and 0.19, respectively.

\end{abstract}


\section{Introduction} \label{sec:intro}
We have surveyed and analyzed the properties of \HI\ line emission in an intermediate latitude field in Ursa Major ($(\ell,\, b) = (143\fdg 6,\, 40\fdg 1)$ or $(\alpha,\, \delta) = (09^{{\mathrm h}}$41$^{{\mathrm m}},\, 68\degree 33')$), focusing on thermal condensation of warm neutral medium gas (WNM) to cold neutral medium gas (CNM) in the intermediate velocity component (IVC).  
This IVC gas is part of the Low-Latitude Intermediate-Velocity Arch (LLIV) studied by \citet{kuntz_danly_1996}, in particular substructure LLIV1 (see their figure 2). 
These substructures (or ``clumps") were cataloged by \citet{kuntz_danly_1996} based on high \HI\ column density (\nh) contours in the IVC range from data from the Bell Laboratories \HI\ survey at 2\degree\ resolution \citep{stark_1992}. LLIV1 has an approximate size of about 10\degree\ on the sky and is connected to other substructures through more extended gas with lower \nh\ (about $10^{19}$\,cm$^{-2}$ in the velocity range $-60$\,\kms\ $< v < -30$\,\kms).
Higher resolution \nh\ maps of LLIV substructures are shown in figures 12(b) and 16 of \citet{wakker_2001}, in the velocity range $-60$\,\kms\ $< v < -30$\,\kms\ and $-70$\,\kms\ $< v < -30$\,\kms, respectively,

Abundances measurements by \citet{wakker_2001} using lines of SII, NI, and OI indicate a metallicity that is approximately solar. Absorption of Fe and Si against the AGN PG 0804+761 at  $(\ell,\, b) = (138\fdg3,\, 31\fdg0)$ indicates some depletion onto dust grains \citep{richterPGabs2001}.
There is thermal dust emission morphologically correlated with the IVC gas in LLIV \citep{planck2011-7.12}.
\citet{wakker_2001} constrain the distance to be in the range 0.9--1.8 kpc (z = 0.6--1.2 kpc). The implied mass is $1.5-6\times10^5\,M_{\sun}$. 

\citet{richter_2003} report a relatively high detection rate of H$_2$ in the IVC gas, implying that CNM is ubiquitous, consistent with a key finding of our paper. 
Absorption lines toward PG 0804+761 \citep{richterPGabs2001} located in the more diffuse part of LLIV at lower Galactic latitudes not analyzed here \citep[][ see their figure 1]{richterPGabs2001} has an estimated H$_2$ mass fraction of $log\,f_{\rm H_2}=-4.5$.
Absorption lines along the line of sight to the distant hot subdwarf PG 0832+675 \citep{richter_2003} characterize the outer part of the substructure LLIV1 studied here; the fraction of H$_2$ is somewhat higher ($log\,f_{\rm H_2}=-3.9$).

Absorption lines toward PG 0804+761 \citep{richterPGabs2001} also indicate a substantial ionization fraction for hydrogen, about 20\%.
There is also hot coronal gas revealed by O VI absorption toward PG 0804+761 \citep{richterPGabs2001} that could potentially be associated with the highly ionized gas related to the envelope of the LLIV Arch. However, this can also be explained by the presence of hot gas above the Perseus arm, a scenario favored by \citet{richterPGabs2001} due to the presence a Galactic chimney found by \citet{normandeau_1996} in the region where the PG 0804+761 line of sight crosses above the Perseus arm. C IV absorption was also detected toward SN 1993J \citep{deBoer1993}.

\citet{wakker_2001} notes how this is reminiscent of a Galactic fountain \citep{shap1976,breg1980,houck1990}, where gas ejected into the Galactic Halo from
inside the Solar circle is expected to have a metallicity slightly above that in the local ISM, while the return flow is outside the Solar circle.  \citet{planck2011-7.12} show that dust appears to survive the hot phases of the flow (or to reform) and discuss the evolution of dust via shattering.  What we find most interesting is that there is a substantial CNM apparently organized with the flow.

Some key elements of the structure of the paper are as follows.
In Section~\ref{sec:data} we present the \HI\ data used in this work.  
Evidence for cold gas in LLIV1 based on \HI\ absorption is summarized in Section \ref{sec:psep}.
Section~\ref{sec:methods} describes the Gaussian decomposition performed to model the \HI\ spectra, Sections \ref{subsec:gh} and \ref{sec:umcomparison} for GHIGLS and DHIGLS, respectively.
Appendices \ref{app:maps} and \ref{app:mapsdh} present maps (2D spatial fields) characterizing each Gaussian component (column density, central velocity, velocity dispersion) for the respective data.
Identification of the different thermal phases is addressed
in Section \ref{subsec:phaseid},
with attention to the robustness of the solution (Section \ref{sec:uncertainties}).
%
Section \ref{sec:gpower} presents an angular power spectrum analysis of the phase maps (Section \ref{subsec:phase}).
The cold gas mass fraction map inferred from the decomposition of DHIGLS data is analyzed in Section~\ref{sec:id-phases}.
Finally, a summary is provided in Section \ref{sec:summary}.

\section{\HI\ Data} \label{sec:data}

\subsection{GHIGLS (9\farcm 4)}
\label{subsec:data-dhigls}

\begin{figure}
    \centering
    \includegraphics[width=\linewidth]{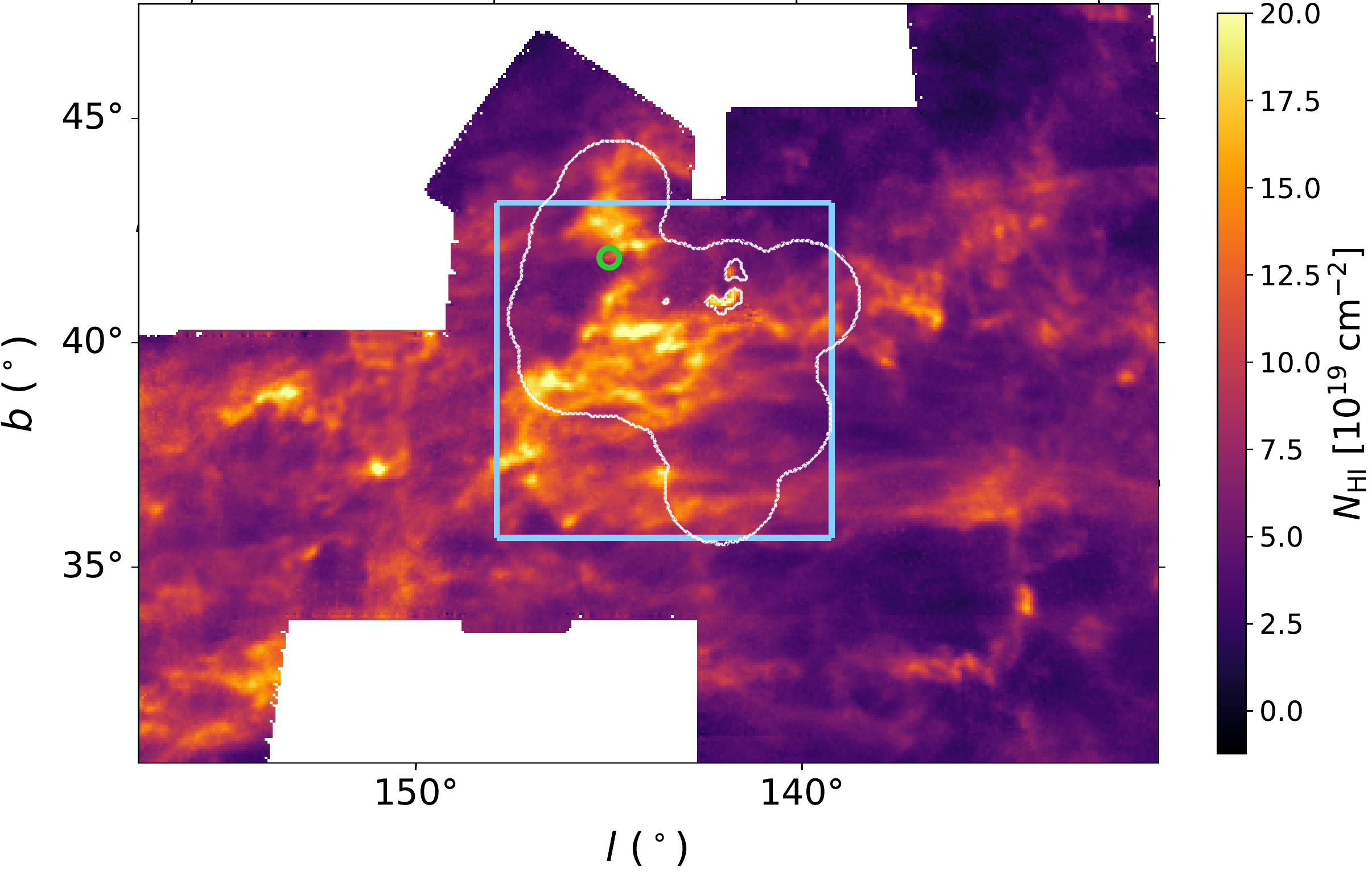}
    \caption{Integrated column density map of the GHIGLS NCPL mosaic in the IVC velocity range  $-81 \leq v \leq -27$\ \kms\ showing the LLIV Arch. The light blue box shows the 128$\times$128 pixel region analyzed. The white contours outline the DHIGLS UM field and its masks (Section \ref{subsec:data-dhigls}). The green ring marks the direction of the absorption measurement (Section \ref{sec:psep}).
    }
    \label{fig:T_b_full_cube_map}
\end{figure}

\begin{figure}
    \begin{center}
        \includegraphics[width=\linewidth]{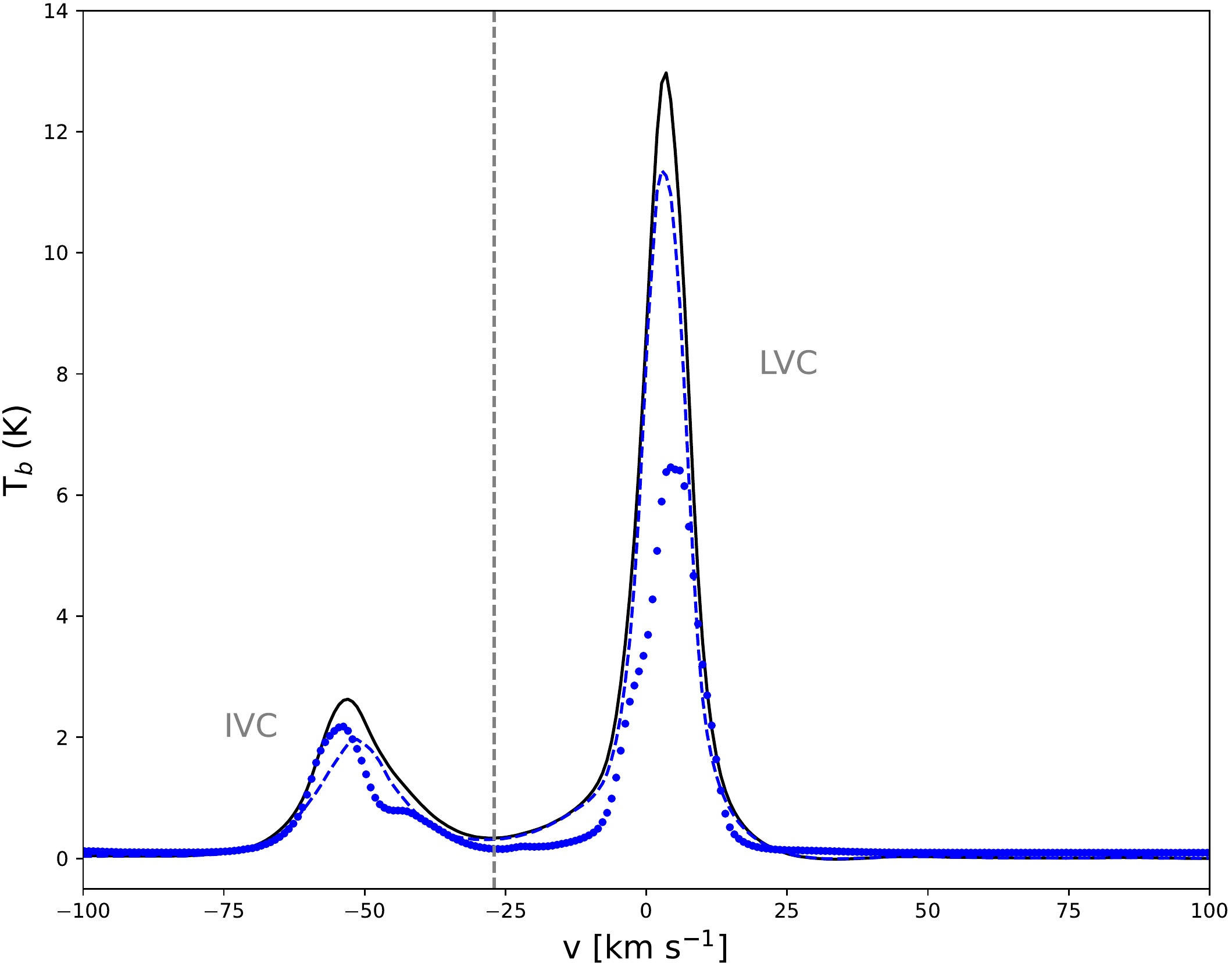}
        \caption{
        GHIGLS \HI\ spectra characterizing the region analyzed in Ursa Major, the light blue box in Figure~\ref{fig:T_b_full_cube_map}, showing the LLIV1 and the LVC gas. The black solid, blue dashed and blue dotted lines are the mean, median, and  standard deviation spectra, respectively.
        } 
        \label{fig:avspec}
    \end{center}
\end{figure}

We used the GHIGLS\footnote{GBT \HI\ Intermediate
Galactic Latitude Survey: \url{https://www.cita.utoronto.ca/GHIGLS/}} 21\,cm line survey \citep{martin_2015} with spatial resolution about  9\farcm4 (pixel size 3\farcm5) and velocity resolution and channel spacing  1.0 \kms \ and 0.8 \kms, respectively, from the Green Bank Telescope (GBT) to examine the brightness temperature $T_b$ of the atomic gas in LLIV1. 
Specifically, we extracted a $7\fdg5$
(128 pixels) 
square sub-region of the NCPL mosaic centered on Galactic coordinates ($144 ^\circ, \ +39^\circ$), the light blue box in Figure \ref{fig:T_b_full_cube_map}. The map shows the total integrated column density of the NCPL mosaic in the velocity range $-81 \leq v \leq -27$\ \kms\ analyzed in this work, covering LLIV1.
Following \citet{boothroyd2011}, the rms error in this column density from the 3D line noise (see Section \ref{subsec:rohsa}) is about $0.14\, 10^{19}$ cm$^{-2}$ for most of the data analyzed, quite low compared to the signal.\footnote{Because of duplicate coverage provided by the fields making up the NCPL mosaic, the noise is lower by about a factor $\sqrt 2$ in the upper left corner and along the right quarter of the field (see the boxy outline of the mosaic in Figure \ref{fig:T_b_full_cube_map} for guidance).} Systematic effects from baseline and stray radiation correction errors could increase this by a factor of three, but these errors do not fluctuate from pixel to pixel and cause the map to look noisy.

Figure~\ref{fig:avspec} shows the mean, median, and standard deviation spectra of data within this sub-region.%
There is a strong peak near 0 \kms \ (hereafter Low Velocity Component (LVC)) that corresponds to gas associated with the NCPL \citep{taank2022}. The secondary peak near $-55$ \kms \ is the LLIV1 gas.
There is also weak emission in the bridge range between the two components.  The ranges were divided at $v=-27$\,\kms\ (denoted by the vertical line) where the emission in the bridge is minimal. 
Individual channel maps of the mosaic show weak emission in the high velocity range from High Velocity Cloud (HVC) complexes A and C \citep{planck2011-7.12}, extending to velocities close to the IVC range (i.e., $\approx-70$\,\kms), but there is very little HVC emission in the region that we analyzed.
For the analysis below, pixels contaminated by extra-galactic emission (mainly the M81/82 group) and masked in GHIGLS because of corrupted baselines are in-painted as described in \citet{taank2022}, appendix A.

\subsection{DHIGLS (1$^\prime$)}\label{subsec:data-dhigls}

\begin{figure}[!t]
    \centering
    \includegraphics[width=\linewidth]{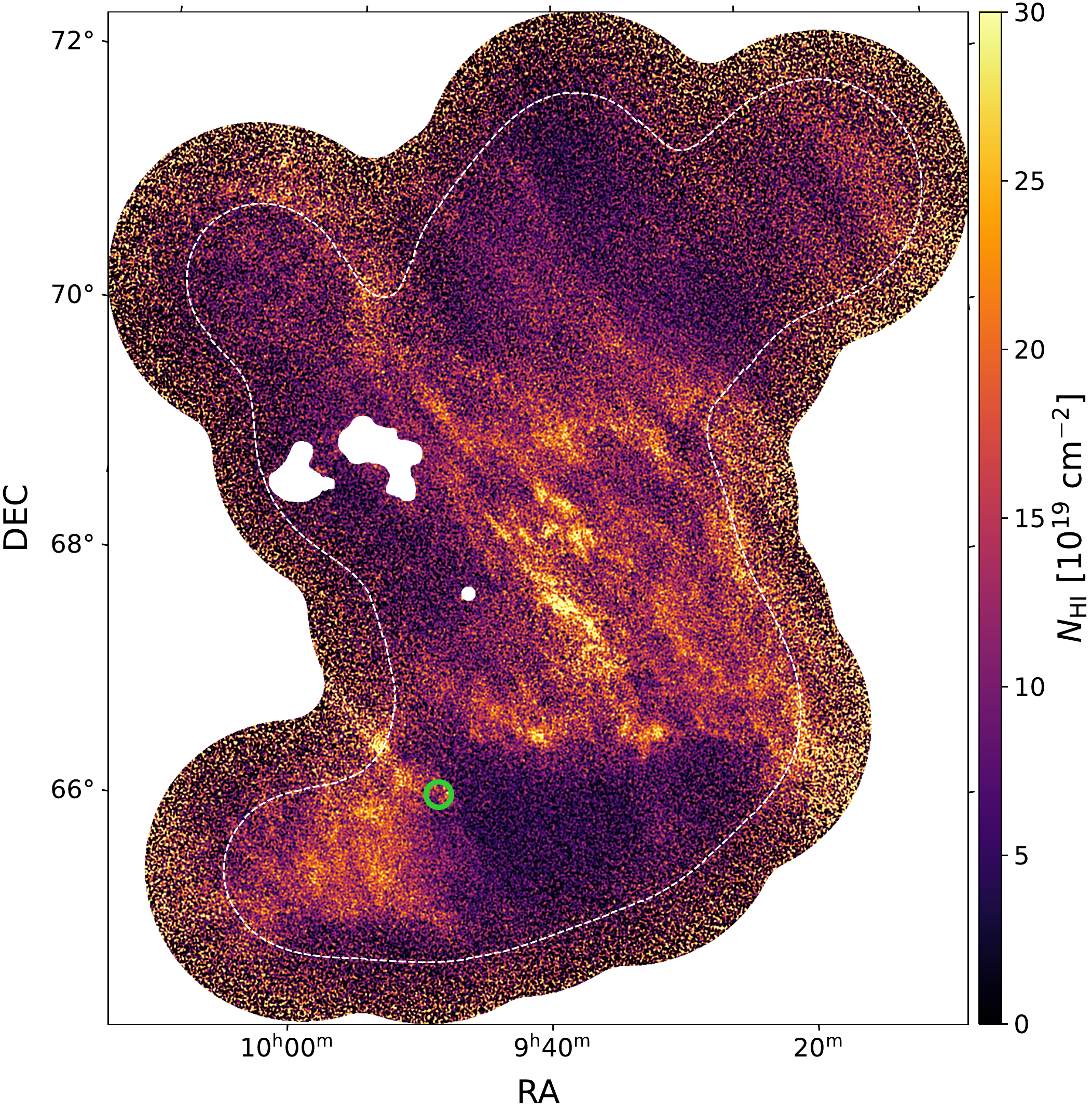}
    \caption{
    Integrated column density map of the IVC gas in UM from DHIGLS. 
    Given the coverage of ST pointings making up the final UM mosaic, the noise is at a minimum toward the center and increases outward, up by a factor of two at the white dashed contour (see text).
    %
    The three white regions inside show the masks applied to avoid extragalactic emission.%
    The green ring marks the direction of the absorption measurement (Section \ref{sec:psep}).
    }
    \label{fig:NHI_UM_IVC}
\end{figure}

We also made use of the 58 square degrees UM dataset that was part of the DHIGLS\footnote{DRAO \HI\ Intermediate Galactic Latitude Survey: \url{https://www.cita.utoronto.ca/DHIGLS/}} \HI\ survey \citep{blagrave_dhigls:_2017} with the Synthesis Telescope (ST) at the Dominion Radio Astrophysical Observatory. This field is located at $(\alpha,\, \delta) = (09^{{\mathrm h}}$41$^{{\mathrm m}},\, 68\degree 33')$ or $(l,\, b) = (143\fdg 6,\, 40\fdg 1)$.  
%
The spectra, with channel spacing $\Delta v=0.824$\,\kms and velocity resolution 1.32\,\kms, cover the range $-70$\,\kms\ $< v < +30$\,\kms.
The spatial resolution of the ST interferometric data was about $1^\prime$.
UM is embedded in the NPCL mosaic from GHIGLS shown in Figure~\ref{fig:T_b_full_cube_map}. 
The DHIGLS UM $T_b$ data cube has the full range of spatial frequencies, obtained by a rigorous combination of the ST interferometric and GBT single dish data \citep[see section~5 in][]{blagrave_dhigls:_2017}. The pixel size is 18\arcsec.

Figure~\ref{fig:NHI_UM_IVC} presents the total column density map of the IVC gas in UM (i.e., LLIV1).  In this DHIGLS data, the error in brightness temperature (which propagates into the column density error) varies with position, increasing at the edge of the map, tracing back to the primary beam response of the antennae in the ST and the (irregular) geometric arrangement of pointings in the full survey \citep{blagrave_dhigls:_2017}. The white dashed contour shows where the noise has increased by a factor of two relative to the minimum of $1.73$\,K in a single channel map or about $1.9\, 10^{19}$ cm$^{-2}$ in the IVC \nh. The full DHIGLS spatial coverage shown here is annotated by the solid white contour in Figure~\ref{fig:T_b_full_cube_map}.
Note the different native coordinate systems: Equatorial for the DHIGLS data and Galactic for the GHIGLS data.
Unlike with the GBT, the higher angular resolution of the ST (i.e., $1^\prime$) allows the detection of high continuum brightness temperature radio point sources, which are useful for \HI\ absorption measurements.

\begin{deluxetable*}{lcccccccccccc}
\tabletypesize{\footnotesize}
\tablecolumns{12} 
\tablewidth{0pt}
\tablecaption{Parameters\tablenotemark{a} of the absorption feature in LLIV1 against 4C~+66.09, interpolated emission, and derived $\Trs$, $f$, and  $M_t$, from DHIGLS data}
\label{table:abs}
\tablehead{
\colhead{Method}  & \colhead{$\Trc$} & \colhead{$\mu_{\mathrm n}$} & \colhead{$\sigma_{\mathrm n}$} & \colhead{$\Trn$} & \colhead{$\mu_{\mathrm b}$} & \colhead{$\sigma_{\rm b}$} & \colhead{$\Trb$} & \colhead{$\Trs$} & \colhead{$\sigma_{\rm th}$} & \colhead{$\sigma_{\rm nt}$} & \colhead{$f$} & \colhead{$M_t$}
}
\startdata
1 & 444 & $2.43$  & $1.52$ & $-109$ & $2.67$ & 1.77 & $22.41$ & 76.3 & 0.80 & 1.31 & 0.51 & 3.3  \\
2 & 444 & $2.43$  & $1.52$ & $-109$ & $2.66$ & 1.99 & $22.73$ & 72.2 & 0.80 & 1.30 & 0.55 & 3.4  \\
\enddata
\tablenotetext{a}{Velocities in \kms\ and temperatures in K.
}
\end{deluxetable*}

\section{Evidence for cold gas from \HI\ absorption in LLIV1}
\label{sec:psep}

\citet{blagrave_dhigls:_2017} reported the detection of CNM gas in the IVC with spin temperature $\Trs=85$\,K against the background radio galaxy ${\rm 4C~+66.09}$ (NVSS\footnote{NRAO VLA Sky Survey:\ \url{https://www.cv.nrao.edu/nvss}, \citet{NVSS-1998}.} ${\rm J094912+661459}$) with a continuum brightness temperature $\Trc=444$\,K. This gas is located within LLIV1 as annotated in Figure~\ref{fig:T_b_full_cube_map}
and related figures.

Following \cite{taank2022}, we used the spectral decomposition code \ROHSA\ (Section \ref{subsec:rohsa}) on the original DHIGLS data to model the emission-absorption pair with Gaussians to refine estimates of the spin temperature and the cold component amplitude and turbulent properties.

We found that the absorption spectrum is well described by a single Gaussian component together with a quadratic polynomial to describe the local residual baseline.
Parameters $\mu_{\mathrm n}$, $\sigma_{\mathrm n}$, and $\Trn$ of the Gaussian are tabulated in Table~\ref{table:abs}.

The associated emission line was obtained by averaging data in an annulus centered on the source with an inner radius $r_{\rm in} = 1\farcm2$ and an outer radius $r_{\rm out} = 3'$ (i.e., 4 and 10 pixels of size 18\arcsec, respectively).
The resulting spectrum is well fit by the sum of two Gaussians, a narrow and a broad component. 
Parameters $\mu_{\mathrm b}$, $\sigma_{\mathrm b}$, and $\Trb$ of the narrow Gaussian are tabulated in Table~\ref{table:abs} (method 1). 
The CNM mass fraction is
$f=0.51$.
Following \citet{blagrave_dhigls:_2017}, the numbers from Table~\ref{table:abs} lead to a spin temperature of $\Trs = 76.3$\,K, slightly lower than their estimate without profile fitting. 
Alternatively, we used \ROHSA\ for a decomposition of emission in a $64\times64$ pixel grid centered on the source. Again, two Gaussians were needed to fully encode the signal. 
Parameters $\mu_{\mathrm b}$, $\sigma_{\mathrm b}$, and $\Trb$ of the narrower Gaussian at the position of the source were interpolated from the {\tt ROHSA} parameter maps in the same annulus and are also tabulated in Table~\ref{table:abs} (method 2). The CNM mass fraction is $f=0.55$ and the spin temperature evaluates to $\Trs=$\,72.2.

Following the procedure described in \citet[][see their section~5.1.2]{taank2022}, for the CNM line we separated the thermal and non-thermal broadening, $\sigma_{\rm th}$ and $\sigma_{\rm nt}$, respectively, and calculated the associated turbulent Mach number $M_t$. These are also tabulated in Table~\ref{table:abs} for methods 1 and 2.

The inferred properties of cold \HI\ gas in LLIV1 confirms the existence of cool atomic gas clumps in the LLIV Arch that was predicted by \citet{richterPGabs2001} based on the H$_2$ excitation temperature $T_{0,1}=193^{+322}_{-75}$\,K found along the line of sight to PG 0804+761. They speculated that $T_{0,1}$ might be representative of the gas kinetic temperature because of thermalization by collisions in a dense cold gas predominantly in the atomic phase (i.e., CNM), the fraction of gas in the molecular phase being low.

\section{Spectral decomposition} 
\label{sec:methods}

To go beyond this single measurement and map out the distribution of cold gas in part of LLIV, specifically LLIV1, we have decomposed emission line data at high resolution from GHIGLS and DHIGLS.

\subsection{\ROHSA}
\label{subsec:rohsa}

\ROHSA\ is a regularized optimization algorithm that decomposes position-position-velocity (PPV) data cubes into a sum of Gaussians \citep{marchal_2019}. {\tt ROHSA} takes into account the spatial coherence of the emission and its multi-phase nature to perform a separation of different thermal phases.
The methodology used in this work is similar to that used in \cite{taank2022}. We refer the reader to their section 3 for a comprehensive description of {\tt ROHSA} and its user-parameters, including the number of Gaussians $N$ and the set of hyper-parameters ($\lambda_{\ab},\lambda_{\mub},\lambda_{\sigmab},\lambda''_{\sigmab}$).
A \ROHSA\ decomposition requires the user to choose this set of parameters to obtain a practicable solution and this must be revisited for a given data set \citep[e.g.,][]{MMG_2021,taank2022}.
Also needed is a noise prescription for the data. For GHIGLS data, we  adopted the 3D prescription discussed by \cite{boothroyd2011}, $S(v,\rb) = S_{e}(\rb)\, \big(1 + \Trb(\rb)/\Tsys\big)$, 
where the 2D map of the standard deviation of the noise $S_e(\rb)$ is  calculated from emission-free end channels (in the case of GHIGLS, supplied with the archival data), and $\Tsys$ is the system temperature, typically 20\,K for the GBT L-band observations. 
We used the augmented version of the \ROHSA\ code employed by \cite{taank2022} to work with 3D noise, rather than the standard 2D noise.
For the DHIGLS data, we used the original implementation of {\tt ROHSA} that considers $S_{e}(\rb)$.

\subsection{GHIGLS}
\label{subsec:gh}

\subsubsection{User parameters} 
\label{subsec:nhsg}

The representative solution presented in this work was obtained using the set of parameters $N = 6$,  $\lambda_{\ab}=\lambda_{\mub}=\lambda_{\sigmab}=40$,  and $\lambda_{\sigmab}^{''}= 50$.

To obtain a solution that fully describes the signal in the IVC range (i.e., LLIV1) without over-fitting the data, we found that $N = 6$ was optimal.
This was accomplished by decomposing the data with varying $N$ in the range $N=[5-9]$ and we used the per-channel mean contribution to chi-square,
\begin{equation}
    \big<\chi^{2}(v)\big> = \sum_{\rb}\left(\frac{{L\big(v, \thetab(\rb)\big)}}{S(v,\rb)}\right)^2\, /\ 128^{2} \, ,
\label{eq:chi-sq-contrib}
\end{equation}
to determine the goodness of fit, where
$L\big(v, \thetab(\rb)\big)$ is the residual between the Gaussian model and the data.
Here the fields $\thetab_n(\rb)=\big(\ab_n(\rb) , \mub_n(\rb) , \sigmab_n(\rb) \big)$ -- amplitude $\ab_{n} \geq \bm{0}$, mean velocity $\mub_{n}$, and standard deviation $\sigmab_{n}$ -- parameterize the $N$ Gaussians of the model prescribed in {\tt ROHSA}.
Note that $128^2$ is the total number of pixels inside the light blue box shown in Figure~\ref{fig:T_b_full_cube_map}.

\begin{figure}
    \begin{center}
        \includegraphics[width=\linewidth]{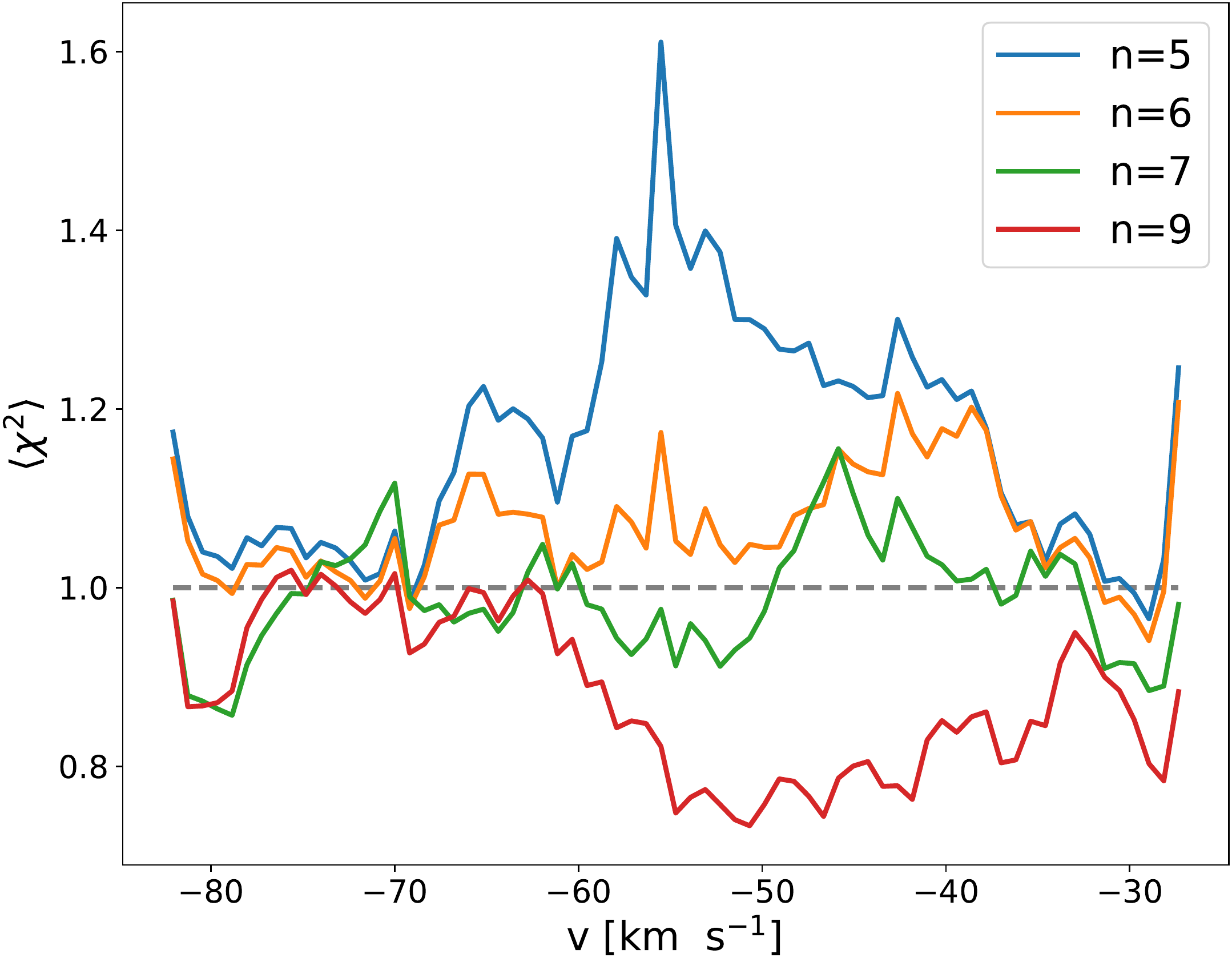}
        \caption{
            Spectrum of mean contribution to chi-square for Gaussian models fit with different numbers of Gaussians.
            Horizontal line indicates $\big<\chi^{2}\big>=1$.
            }
        \label{fig:contchi2comp}
    \end{center}
\end{figure}

\begin{figure}
    \begin{center}
        \includegraphics[width=1.005\linewidth]{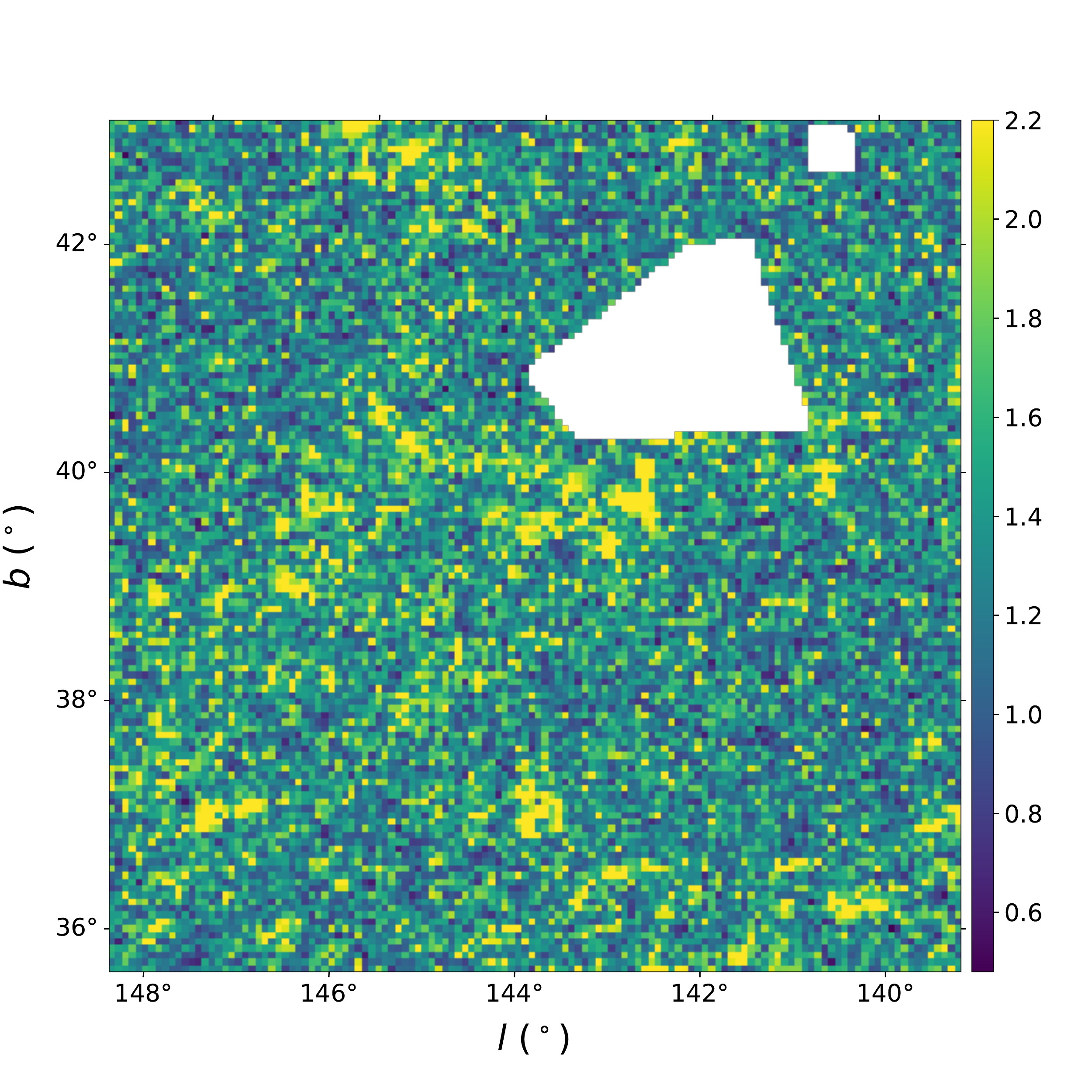}
        \caption{
        Map of the reduced $\chi_r^{2}$ obtained with {\tt ROHSA} on the GHIGLS data of LLIV1.
        Masked regions correspond to pixels contaminated by extragalactic emission and in-painted before decomposition.
    }
        \label{fig:chi2}
    \end{center}
\end{figure}

Figure~\ref{fig:contchi2comp} shows spectra of the mean contribution to chi-square $\big<\chi^{2}(v)\big>$ for varying $N$, where
$N=6$ provides a solution in which $\big<\chi^{2}(v)\big>$ is just slightly higher than unity denoted by the horizontal dashed line. For $N>6$, $\big<\chi^{2}(v)\big>$ is dominated by values lower than unity, especially where the signal is strong, which indicates over-fitting the data.

Figure~\ref{fig:chi2} shows a reduced but spatially resolved version of Equation~\ref{eq:chi-sq-contrib},

\begin{equation}
  \label{eq:chi2}
  \chi^2_r(\thetab(\rb)) = \sum_{v} \left(\frac{L\big(v, \thetab(\rb)\big)}{\bm{S}(v,\rb)}\right)^2 / \, k \, ,
\end{equation}
where $k = 69 - 3 N$ is the number of degrees of freedom, with $69$ being the number of velocity channels.
Spectral data that were in-painted prior to decomposing the cube was not considered and are shown as masked regions.
Our best model achieves a mean $\chi^2_r$ across the field of 1.4, with median and standard deviation of 1.3 and 0.4, respectively.

\begin{figure*}
\centering
\includegraphics[width=0.98\linewidth]{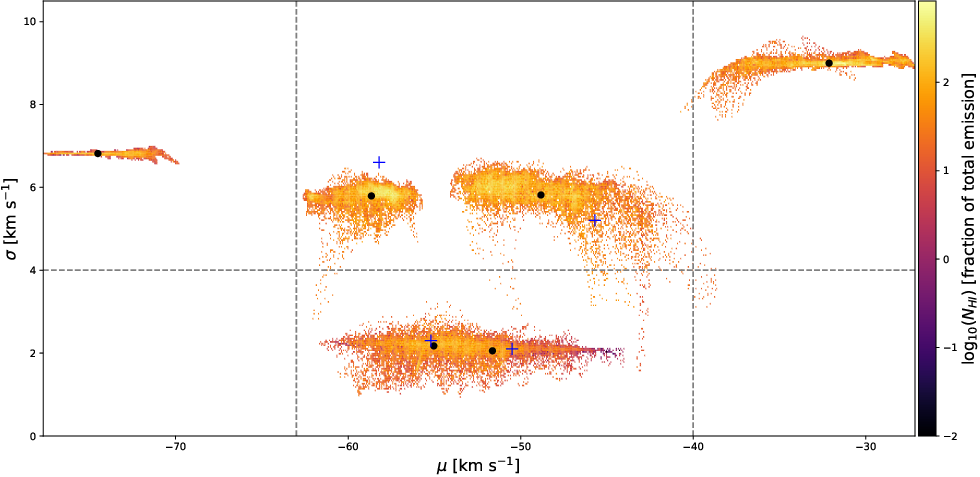}
        \caption{
        2D histogram of $\sigma$ and $\mu$ of the six Gaussians, weighted by their column densities obtained by decomposing the GHIGLS data of LLIV1 with {\tt ROHSA}.
        The black dots correspond to the column density means of each cluster. 
        The blue crosses show the same quantity but for the decomposition of the DHIGLS data within the selected velocity range. 
        The horizontal dotted grey line shows the separation between the identified two thermal phases.  
        The vertical grey lines show the velocity range where Gaussian components encodes emission from LLIV1.
        } 
\label{fig:2d_hist}
\end{figure*}

The four hyper-parameters that control the smoothness of the Gaussian parameter maps were chosen to be the same due to the similar functional form of their respective cost functions. This efficiency limits the parameter space explored in finding a practicable solution. 
They were chosen empirically to correlate adjacent pixels on a spatial scale close to the beam of the instrument. In experimenting with this set of parameters, the optimal model was found to have $\lambda_{\ab}=\lambda_{\mub}=\lambda_{\sigmab}=40$.

To ensure a solution in which phases are identifiable according to distinct velocity dispersions (phase separation), we explored a range of the $\lambda''_{\sigmab}$ parameter, which controls the variance of the normalized velocity dispersion of Gaussian components. We experimented with $\lambda''_{\sigmab}= [1,10,20,50,100,500,1000$] and found that 50 allows a coherent phase separation (see Section~\ref{subsec:phaseid}), without adding too much penalty in the global cost function prescribed in \ROHSA\ that would prevent an overall good fit of the data.

\subsubsection{A representative solution}
\label{subsec:phaseid}

\begin{deluxetable}{lccccccc}
    \tablecaption{Mean kinematic properties (in \kms) of Gaussians encoding the IVC gas (LLIV1) in the GHIGLS and DHIGLS data
    }
    \label{table:mean_var}
    \tablewidth{0pt}
    \tablehead{
    \colhead{Survey} & \nocolhead{}  & \colhead{$G_0$} & \colhead{$G_1$} & \colhead{$G_2$} & \colhead{$G_3$} & \colhead{$G_4$} & \colhead{$G_5$}}
    \startdata
    GHIGLS & $\left<\mub_n\right>$ & $-73.6$ & $-58.7$ & $-55.6$ & $-52.9$ & $-48.7$ & $-33.0$ \\ 
    & $\left<\sigmab_n\right>$ & $6.8$ & $5.7$ & $2.1$ & $2.0$ & $5.8$ & $9.08$  \\ 
    DHIGLS & $\left<\mub_n\right>$ & & $-58.2$ & $-55.2$ & $-50.5$ & $-45.7$ & $-26.5$  \\ 
    & $\left<\sigmab_n\right>$ & & $6.6$ & $2.3$ & $2.1$ & $5.2$ & $10.0$   \\ 
    \enddata
\end{deluxetable}

The derived Gaussian model parameters for all spatial pixels are summarized in the 2D histogram of the column density weighted $\sigma - \mu$ parameters in Figure~\ref{fig:2d_hist}. There are six clusters of points that correspond to the $N=6$ Gaussians used by \ROHSA\ to fit the data. 
Black points show the column density-weighted average of each cluster, as summarized in Table~\ref{table:mean_var}. The distinct average velocity dispersions of the six clusters, ranging from about two to nine \kms, reveal the multiphase nature of the gas in LLIV1.

We observe a clear separation of clusters vertically along the velocity dispersion axis, denoted by the grey horizontal dashed line. $G_0$, $G_1$, $G_4$, and $G_5$ are broad, typical of warm gas, while $G_2$ and $G_3$ are narrow and can be associated with the cold phase of LLIV1.
Interestingly, unlike what was found in other fields -- LVC from GHIGLS in the North Ecliptic Pole  \citep{marchal_2019,marchal_miville_2021} and the NCPL \citep{taank2022}, and HVC in complex C from DHIGLS in EN \citep{MMG_2021} -- there are no Gaussian components with intermediate dispersion characteristic of lukewarm gas (LNM) associated with a thermally unstable medium. 
This suggests that in LLIV1 the thermal condensation and phase transition are not ongoing, but rather have reached an equilibrium state.

Horizontally, along the velocity axis, the components $G_1$ to $G_4$ are located near the peak of LLIV1 emission as shown in Figure~\ref{fig:avspec}. $G_0$ describes gas in the HVC range that is not of interest in this work and $G_5$ gas is located in the velocity bridge between emission from the LVC and LLIV1.

The corresponding Gaussian parameter maps, sorted by increasing mean velocity, are presented in Appendix~\ref{app:maps}. Column density, velocity, and dispersion velocity maps of the six Gaussian components are shown in Figures~\ref{fig:col-dens-mosaic}, \ref{fig:vel-mosaic}, and~\ref{fig:sigma-mosaic}, respectively.
Inspecting each column density map, we found that $G_1$, $G_2$, $G_3$, and $G_4$ show a morphological correlation not shared with $G_0$ or$G_5$.
In the following sections, only the four Gaussian components between the two vertical dashed lines in Figure~\ref{fig:2d_hist} were kept.

\subsubsection{Uncertainties}
\label{sec:uncertainties}

\begin{figure}
    \begin{center}
        \includegraphics[width=\linewidth]{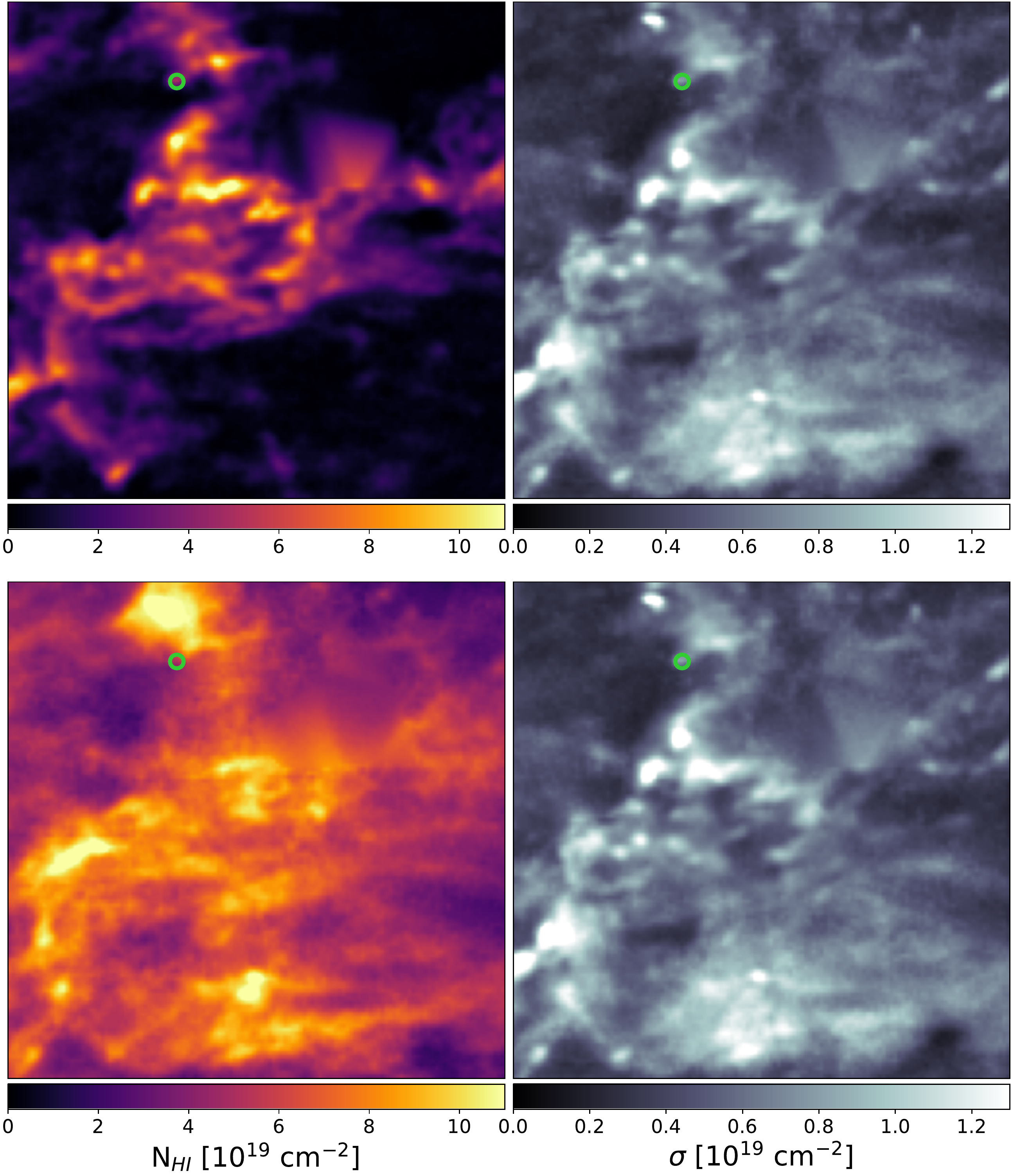}
        \caption{
        Column density maps (first column) with their associated uncertainties  (second column) of the cold (top) and warm (bottom) phases in the GHIGLS data of LLIV1. 
        The green ring marks the position of the absorption feature against 4C~+66.09.
        }
        \label{fig:nhi_maps_ghigls}
    \end{center}
\end{figure}

To explore the degeneracy of the \ROHSA\ solution, we generated a model cube from the mean representative solution, $G_1$ to $G_4$ in Table \ref{table:mean_var},
and repeated the Gaussian decomposition using three series of 50 runs each \citep{MMG_2021,taank2022}. 
The first series explored how the outcome of the decomposition was influenced by the injection of 50 different instances of noise. The hyper-parameters were kept the same. 
The second series explored the impact of the hyper-parameters,  
with $50$ runs using random perturbations of the four \ROHSA\ hyper-parameters in a $\pm10$\% interval around the original values. Here, the injected noise
was kept the same.
The third explored the sensitivity to %
the four Gaussians needed to initialize \ROHSA, with 50 runs selecting them 
randomly from the original $\sigma$ -- $\mu$ diagram in Figure~\ref{fig:2d_hist}. We refer the reader to \citet{taank2022} for further details of that procedure.

For each of the 50-run series, the outcomes were examined in $\sigma$ -- $\mu$ space, revealing that the clusters observed as in Figure~\ref{fig:2d_hist} were quite stable, including the lack of components with intermediate velocity dispersions.
For each run, we performed a phase separation by grouping the four Gaussians into two categories based on their mean velocity dispersion; Gaussians with $\left<\sigmab_n\right>>4$\,\kms\, were classified as WNM and Gaussians with $\left<\sigmab_n\right><4$\,\kms\, were classified as CNM. 

For each series, maps of the mean column density and its standard deviation were generated for the two components.
Results from all 150 runs were combined to calculate the final maps of the column density for the two components. 
For both of these, the standard deviations from the three series were summed in quadrature to yield the total uncertainty.

\subsubsection{Phase maps}
\label{subsec:phase}
Figure~\ref{fig:nhi_maps_ghigls} shows the column density maps (first column) with their associated uncertainty maps (second column) of the cold (top) and warm phases (bottom) in the GHIGLS data of LLIV1.
There are some variations of the uncertainties across the field, but the low values of the uncertainties relative to the associated column densities indicate the good stability of the solution found with \ROHSA.

The cold phase of LLIV1 appears as a collection of elongated filaments that forms a closed structure within the decomposed field. 
These substructures follow the orientation of the overall large-scale cloud, along the diagonal of the field from the northwest to the southeast (Galactic coordinates).
The column density of the more diffuse warm phase is highest within the contour delimited by the presence of cold gas but also exists outside of LLIV1.

\begin{figure}
\centering
\includegraphics[width=\linewidth]{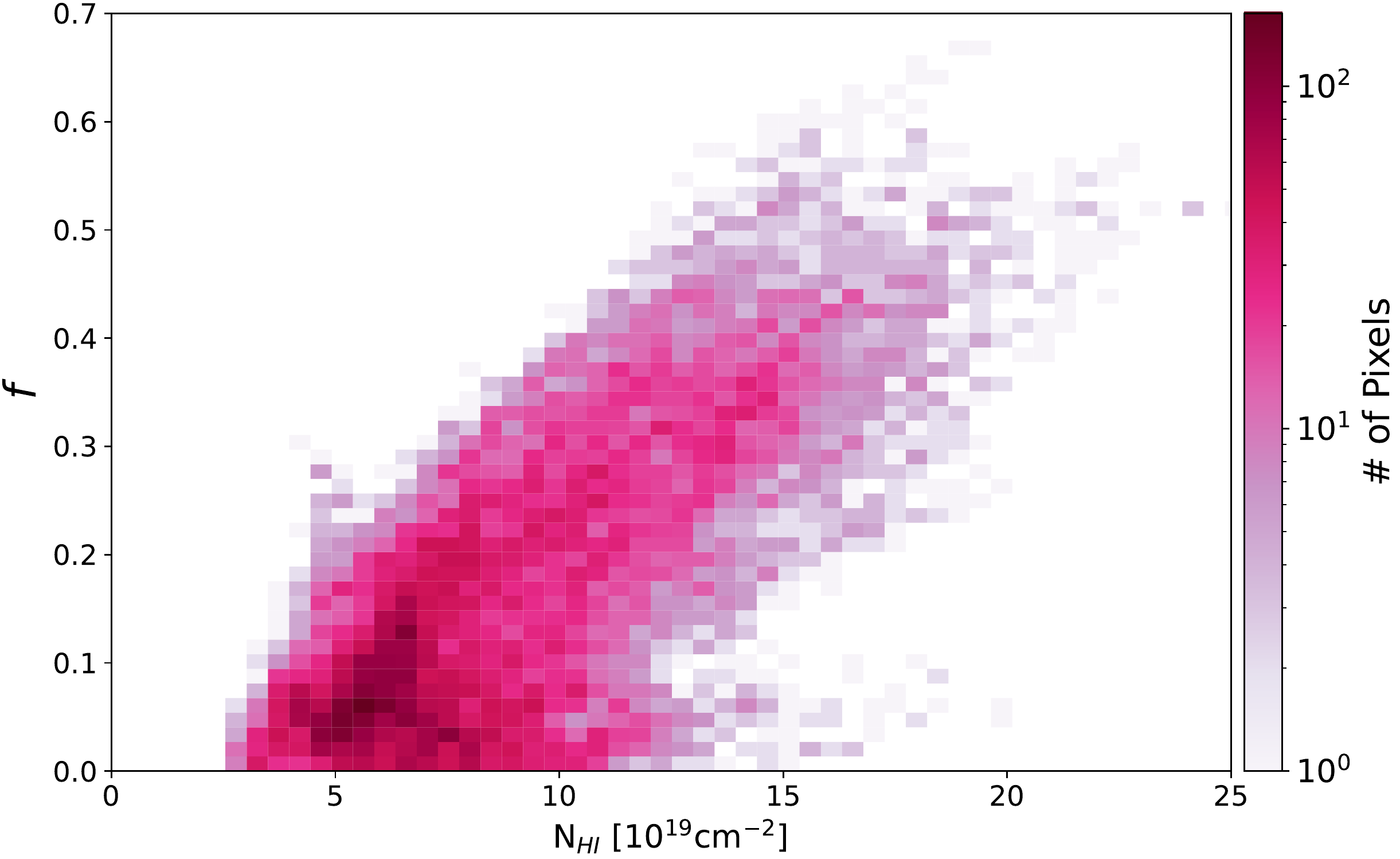}
        \caption{2D histogram of the CNM mass fraction with respect to the total \nh, from the GHIGLS solution.
        }
        \label{fig:massfracscat}
\end{figure}

Statistically, the cold gas mass fraction $f$ increases with the total gas column density, as seen in the 2D histogram in Figure \ref{fig:massfracscat}.\footnote{Note that if $f \propto \nh(\mathrm{total})$ then $\nh(\mathrm{CNM)} \propto \nh(\mathrm{total})^2$, which non-linearity is borne out in the alternative 2D histogram $\nh(\mathrm{CNM)}\,  \mathrm{vs.}\, \nh(\mathrm{total}$ (not shown).} In physical structures like in LLIV1, increases in column density are likely indicative of increased volume density, rather than increased path length along the line of sight. Thus the trend in Figure \ref{fig:massfracscat} suggests that $f$ increases with volume density. In any instability leading to CNM, cooling relative to heating increases most rapidly in dense gas, increasing the efficiency and shortening the timescale of thermal condensation. Perhaps this is an intuitive basis for the trend, but numerical simulations would be needed for a full understanding.
Beyond what is seen statistically in Figure \ref{fig:massfracscat}, the spatial dependence of (high) $f$ will be quantified in Section~\ref{sec:id-phases}.

\subsubsection{Power spectrum}
\label{sec:gpower}

We statistically quantified and compared the multi-scale structure of the two phases by calculating for each the angular power spectrum $P(k)$, the azimuthal average of the modulus of the Fourier transform of the column density field. Following \citet{martin_2015} and \citet{blagrave_dhigls:_2017}, we modelled them as $P(k) = B(k) \times P_0 k^{\gamma} + A \times N(k)$, 
where $P_0$ is the amplitude of the power spectrum, $\gamma$ is the scaling exponent, $B(k)$ describes the cutoff of the spectrum at high $k$ due to the beam of the instrument, assumed to be a 2D Gaussian of FWHM = 9\farcm4, and $N(k)$ is the noise estimated by taking the power spectrum of empty channel maps of the PPV cube and scaled by a multiplicative factor $A$. The finite images were apodized using a cosine function to minimize systematic edge effects in the implementation of the Fourier transform.

\begin{figure}
    \begin{center}
        \includegraphics[width=\linewidth]{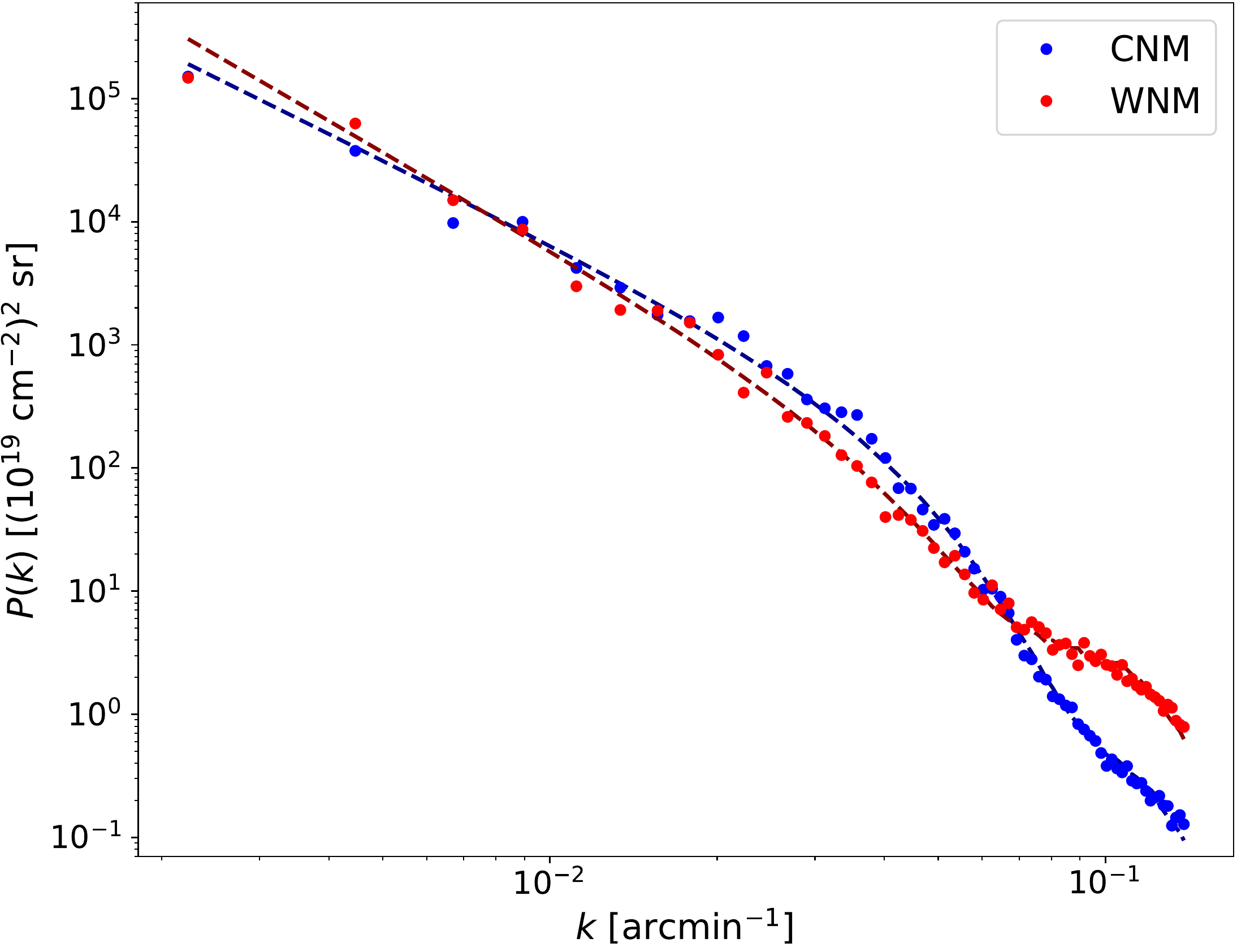}
        \caption{
        Angular power spectra $P(k)$ of the integrated column density of the cold (blue) and warm (red) phases.
        The dashed lines represent the models fit to each phase independently.
        }
        \label{fig:gpowerspec}
    \end{center}
\end{figure}

Figure~\ref{fig:gpowerspec} shows $P(k)$ for the cold and warm phases, in blue and red dots respectively. The total fitted model is shown by the dashed dark blue and red curves for the cold and warm phases, respectively.
Recognizing the uncertainties, we find that the scaling exponent for the cold phase $\gamma_{\rm CNM}=-2.230\pm 0.002$ is higher than that of the warm phase with $\gamma_{\rm WNM}=-2.613\pm 0.003$. In other words, the power spectrum of the cold phase in slightly shallower than that of the warm phase, quantifying that the cold phase has relatively more structure on small scales.
This is similar to what was found in high latitude LVC gas in the NEP field of the GHIGLS survey \citep{marchal_miville_2021} and in  HVC gas (complex C) in the EN field of the DHIGLS survey \citep{MMG_2021}.

\subsection{DHIGLS}
\label{sec:umcomparison}

\begin{figure*}
    \begin{center}
        \includegraphics[width=0.475\linewidth]{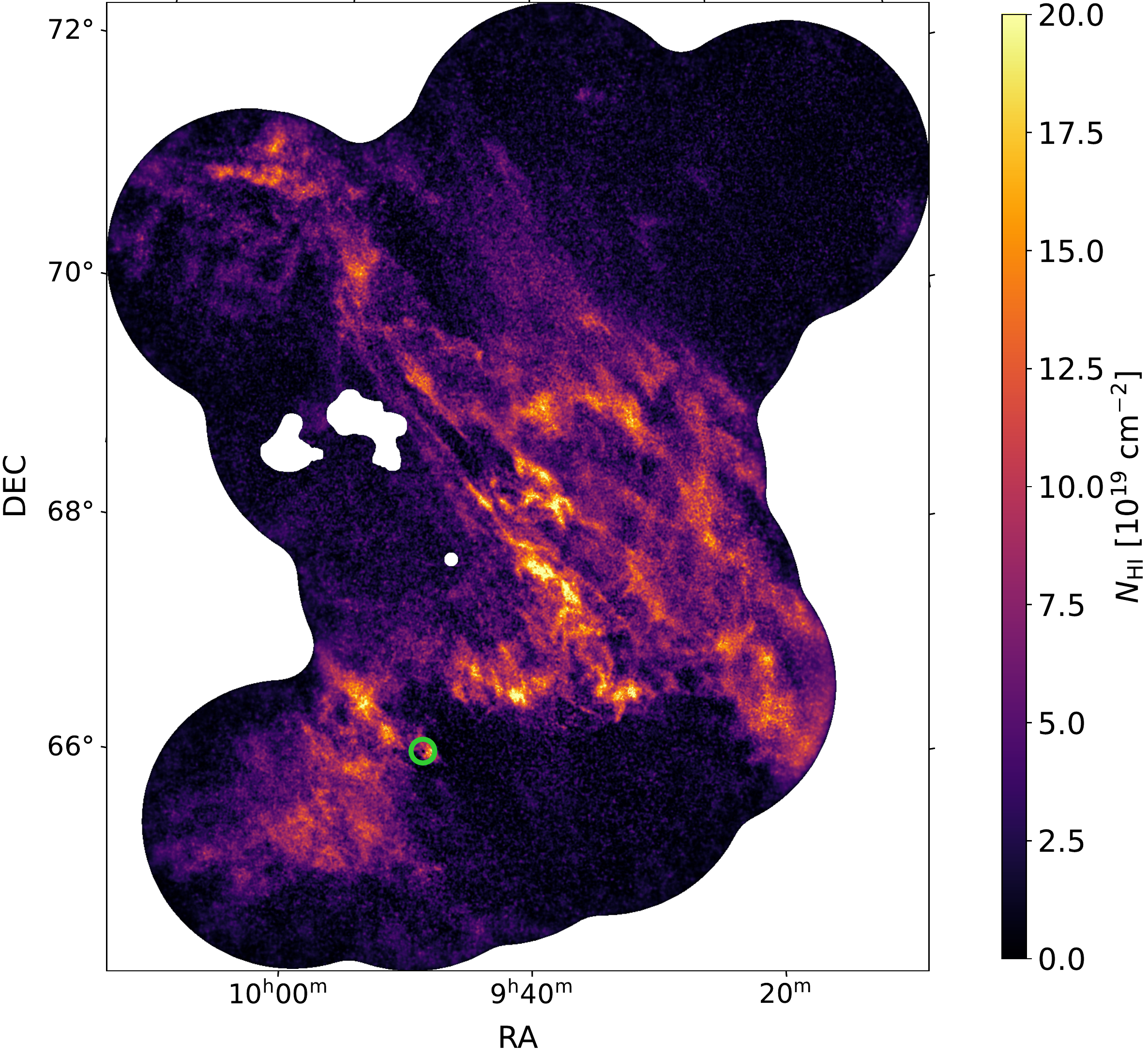}
        \includegraphics[width=0.475\linewidth]{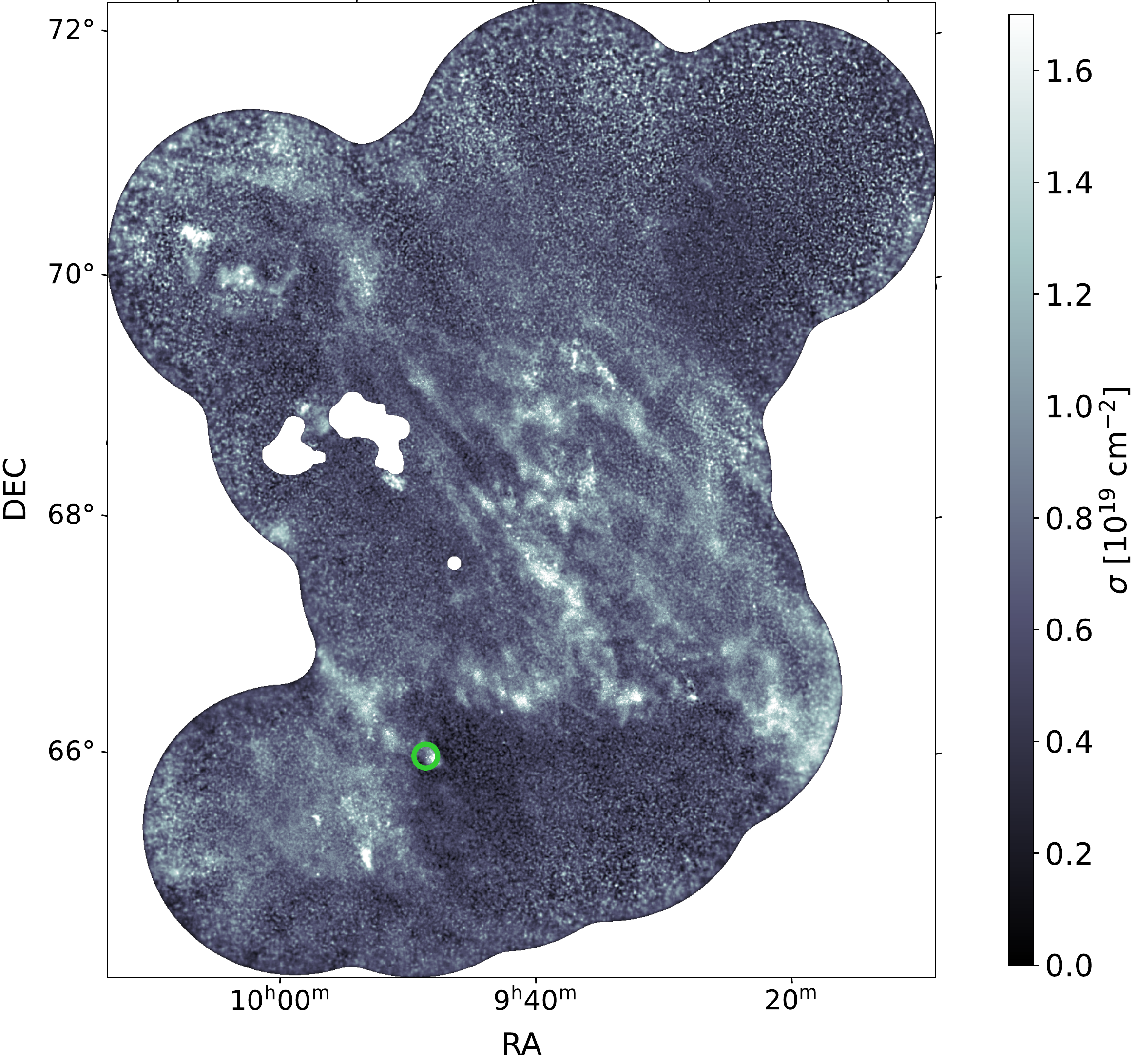}
        \includegraphics[width=0.475\linewidth]{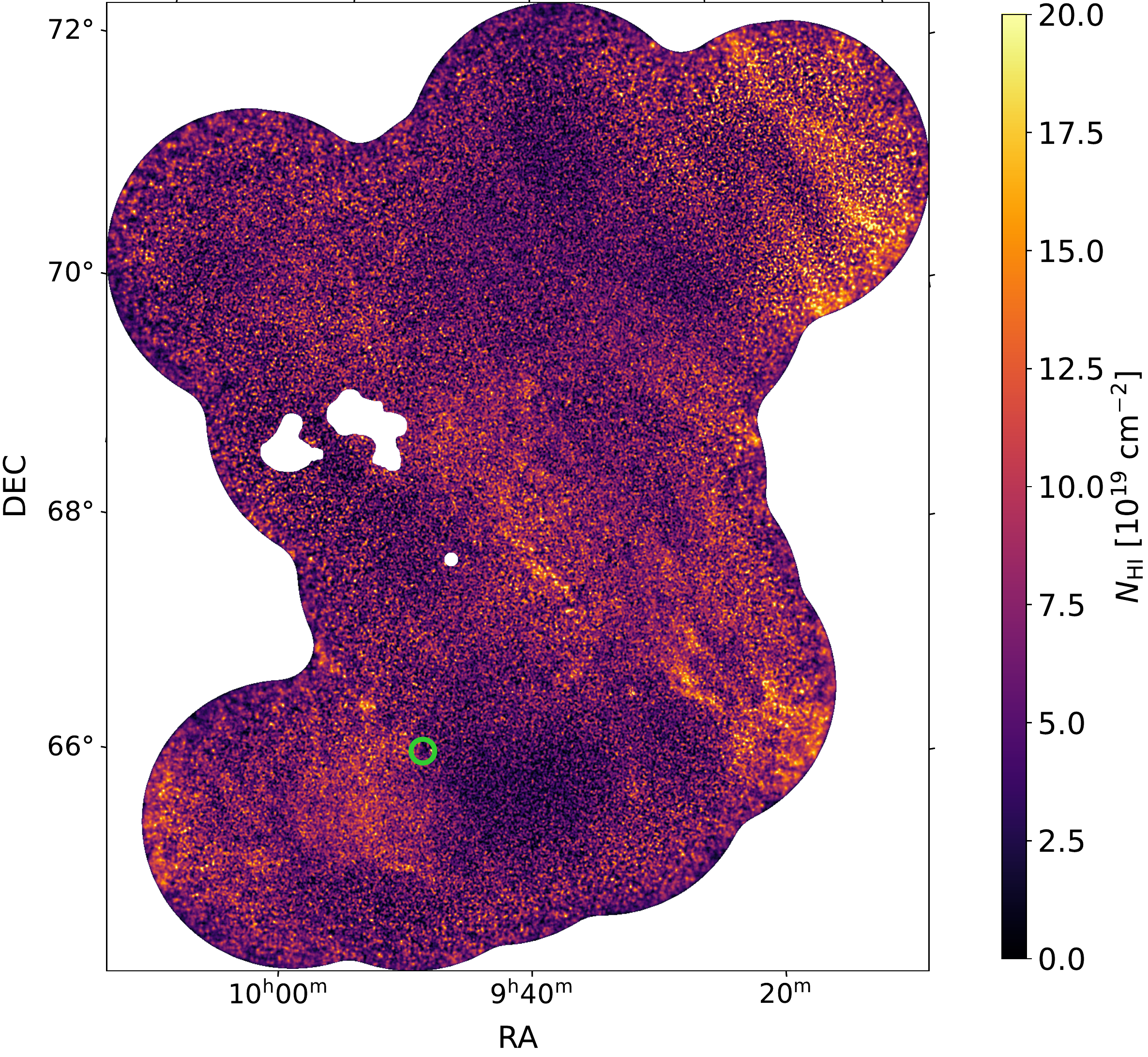}
        \includegraphics[width=0.475\linewidth]{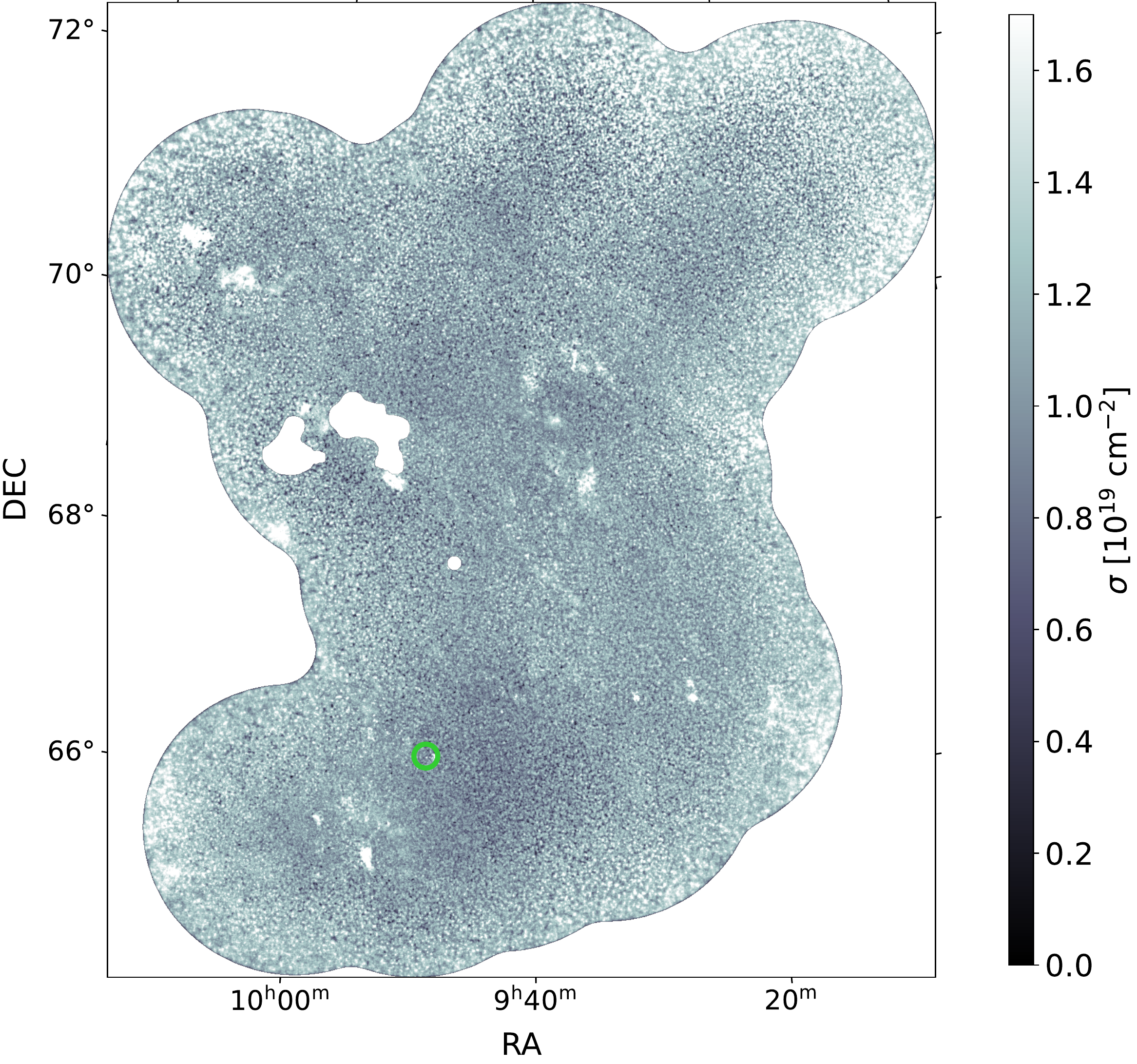}
        \caption{
        Column density maps (first column) with their associated uncertainty maps (second column) of the cold (top) and warm (bottom) phases from the DHIGLS data in LLIV1.
        The green ring marks the position of the absorption measurement against 4C +66.09.
        }
        \label{fig:umargb}
    \end{center}
\end{figure*}

\subsubsection{Initialization}
We performed a first exploration of the user parameters using the same methodology as described in Section \ref{subsec:nhsg}. This search resulted in a set of parameters ($N=5$, $\lambda_{\ab}=\lambda_{\mub}=\lambda_{\sigmab}=10$, and $\lambda_{\sigmab}^{''}= 10$) that could satisfy our selection criteria, including a flat $\chi_r^2$ map with a mean value close to unity.
Although statistically good, this solution was found to be inadequate. After generating the model cube from the solution and applying a convolution to both the model and the data to lower their spatial resolution, the model was no longer providing a good description of the data. 

Specifically, on some lines of sight where the convolved data show two distinct peaks even at a resolution as low as 2', the model solution was found to have only one narrow Gaussian to describe the cold gas. 
Figure~\ref{fig:multirescon} illustrates the spectrum on such a line of sight,  showing the original data (black) and the data convolved to  2' and 9\farcm4 resolution in green and blue, respectively. The improved model, described next, fit to the DHIGLS data is shown by the red curve.

To obtain a consistent solution at various resolution (from the native $1^\prime$ beam to the 9\farcm4 beam of the GHIGLS data), we used the GHIGLS values tabulated in Table~\ref{table:mean_var} as initial parameters to start the multi-resolution process prescribed in \ROHSA\  \citep{marchal_2019,taank2022}.
Only the five components whose mean velocities were included in the velocity range covered by the DHIGLS data where used: Gaussians $G_1$ to $G_5$.

\subsubsection{Solution and Uncertainties}
\label{sec:dphase}

An appropriate set of hyper-parameters was again found to be $\lambda_{\ab}=\lambda_{\mub}=\lambda_{\sigmab}=10$, and $\lambda_{\sigmab}^{''}= 10$.
The resulting decomposition still satisfies all of our selection criteria, with a fairly uniform $\chi_r^2$ map (not shown here) and showing no evidence of over-fitting  the data.

The mean kinematic properties of the five Gaussians used to describe the DHIGLS data are tabulated in Table~\ref{table:mean_var}.
Values of the components $G_1$ to $G_4$ (associated with LLIV1) are also denoted by blue crosses in Figure~\ref{fig:2d_hist}, appearing to be fairly consistent with those obtained from the decomposition of the GHIGLS data.

The corresponding parameter maps of the five Gaussian components, column density, velocity, and velocity dispersion,  sorted by increasing mean velocity, are presented in Appendix~\ref{app:mapsdh}, Figure~\ref{fig:mosaic_params}.
Inspection of each column density map shows similar morphology (and morphological correlation between components) to those of the GHIGLS solution, described in Section~\ref{subsec:phaseid} and Appendix \ref{app:maps}.

Building on this solution, we followed the same methodology as described in Section~\ref{sec:uncertainties} to evaluate the mean column density and uncertainty maps for each phase.
The top and bottom panels in Figure~\ref{fig:umargb} show the resulting maps for the cold and warm phases, respectively. 

\subsubsection{Comparison with GHIGLS}

Our modeling of the cold phase in LLIV1 at high resolution ($1^\prime$) reveals similar large scale properties as the 9\farcm4 modeling obtained with GHIGLS data; the cold phase seems to be confined within a closed contour, surrounded by or mixed in with its warm counterpart.
The global orientation of LLIV1 and the filamentary structures within suggests that there is a bulk motion of the gas cloud on the plane of the sky (from top left to bottom right in Equatorial coordinates).

The cold filamentary structures appear thinner at this resolution and the width of some filaments likely reaches the size of the $1^\prime$ beam of the DHIGLS data.
This is notably the case for the highly elongated filament seen 
just to the right of the masked regions within the field and oriented along its diagonal (from top left to bottom right).
Clustering the filaments and analyzing their statistical properties \citep{MMG_2021} is beyond the scope of this paper but in further work would provide valuable insight into the thermal condensation that has occurred in LLIV1.

\subsubsection{CNM mass fraction}
\label{sec:id-phases}

\begin{figure*}
    \centering
    \includegraphics[width=0.49\linewidth]{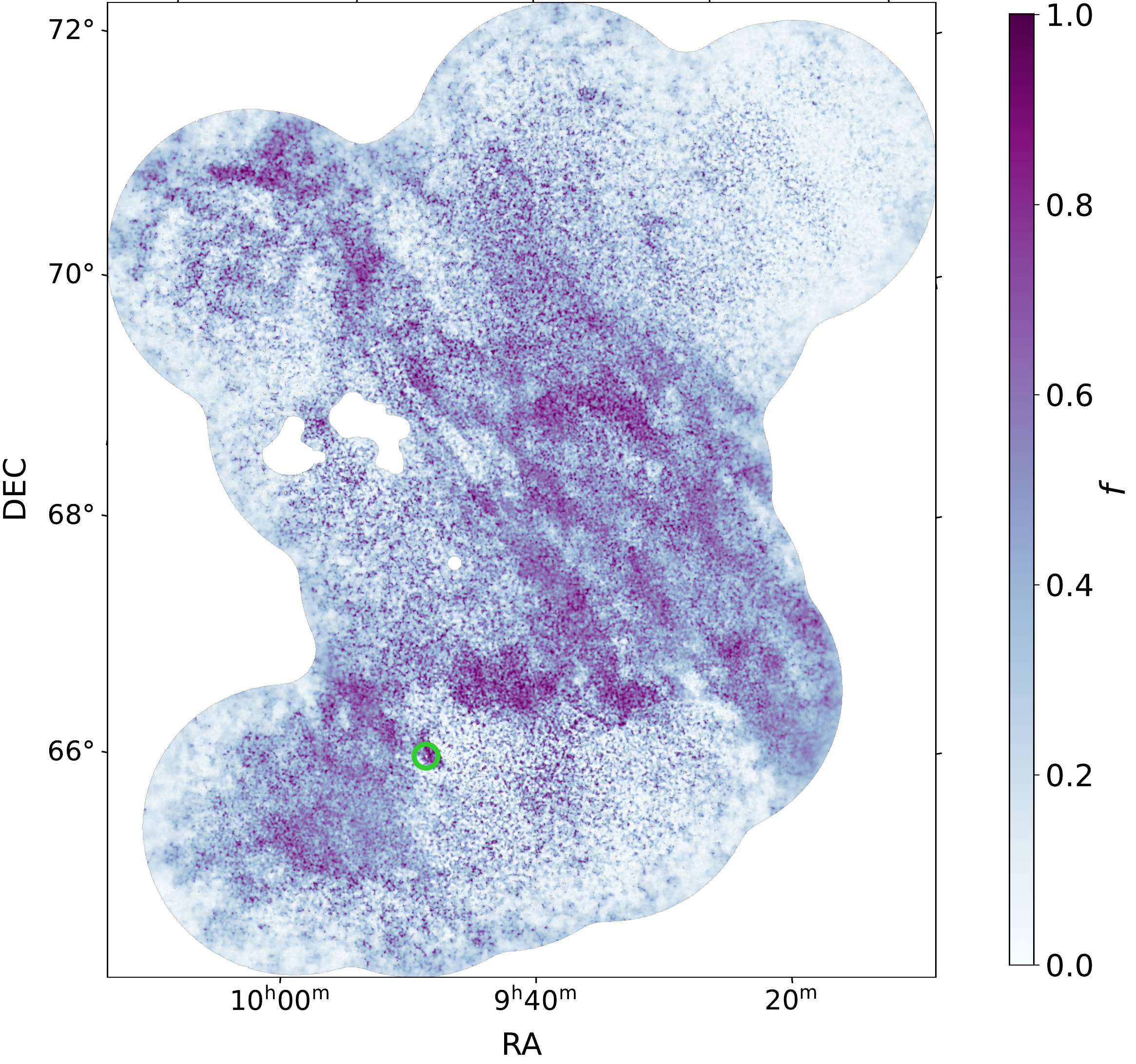}    \includegraphics[width=0.49\linewidth]{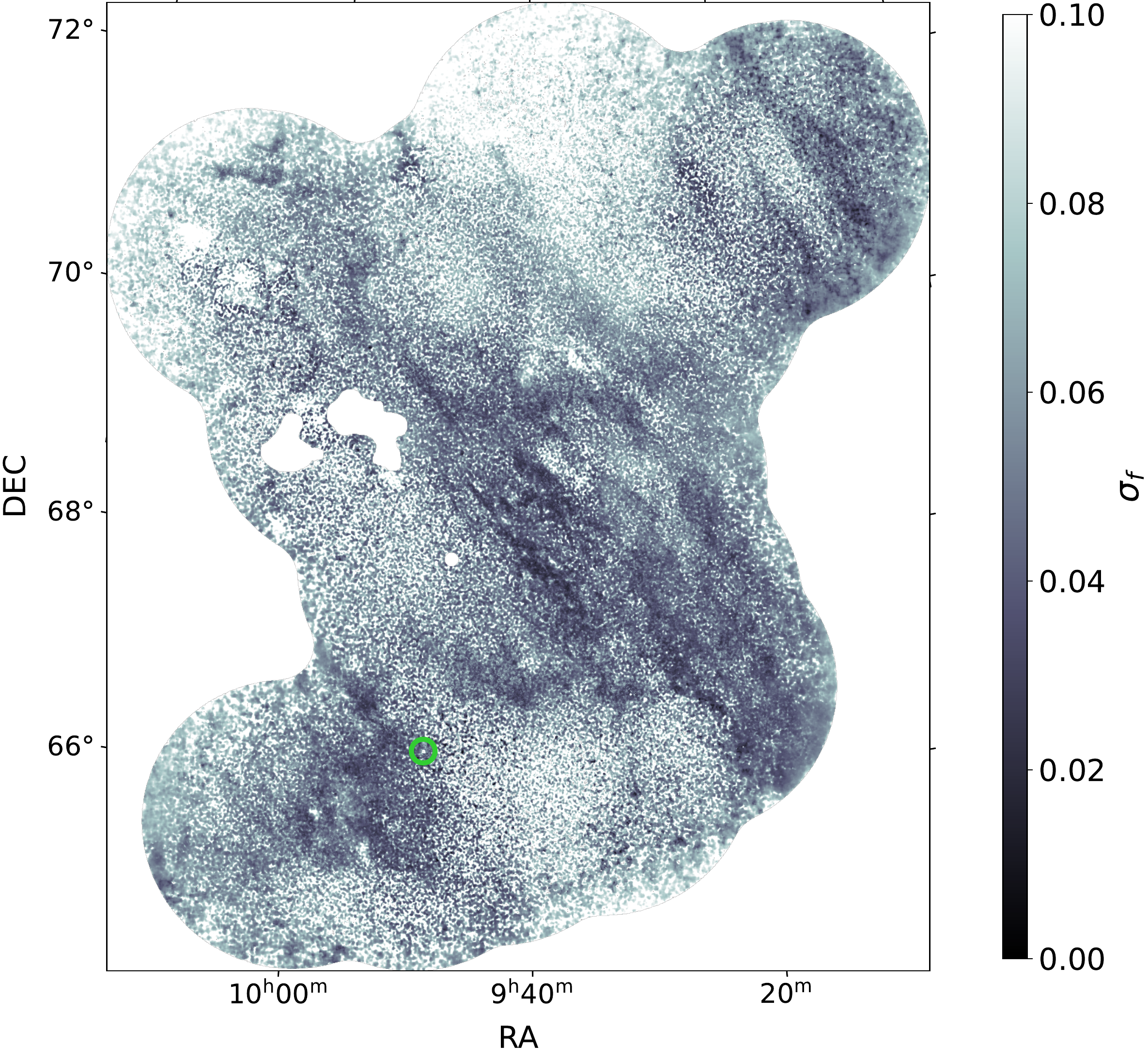}
    \caption{CNM mass fraction map and uncertainty inferred from the DHIGLS data.}
    \label{fig:fcnm_UM_IVC}
\end{figure*}

The left panel in Figure~\ref{fig:fcnm_UM_IVC} shows our best model of the cold gas mass fraction $f(\rb)$ in LLIV1 at the $1^\prime$ resolution of the DHIGLS data, and the left panel shows the corresponding uncertainties.
Large variations of $f(\rb)$ occur 
across the field %
and even within the contour delimiting the large scale extent of the cold phase. 
The mean and standard deviation of $f(\rb)$ are 0.33 and 0.19, respectively.
Interpolated at the position of the absorption measurement against 4C~+66.09, $f(\rb)$  is $0.46\pm 0.06$, close to the values tabulated in Table~\ref{table:abs}.

The spatial distribution (i.e., morphological structure) of $f(\rb)$ resembles the column density map of the cold phase only, due to a relatively constant (i.e., flat) column density observed in the warm phase within the sky coverage of the DHIGLS data.
Specifically, $f(\rb)$ shows the same elongated filaments, as well as finger-like structures that  seem to be part of an organized series of scalloping structures at the implied leading edge of LLIV1, oriented perpendicular to the large-scale orientation of the cloud and large filaments.
This is reminiscent of the structures observed in the Draco Nebula \citep[][and references therein]{miville_2017}, another IVC that is thought to be part of the Galactic fountain process.%

\section{Summary} 
\label{sec:summary}

Our novel study of the multi-phase properties of LLIV1 is based on \HI spectra from GHIGLS and DHIGLS. 
We used \ROHSA\ to decompose the spectral data in emission to model their multiphase structure and corroborated this with analysis of an absorption spectrum. %
The main conclusions are as follows. 

\begin{itemize}
    \itemsep-0.2em
    \item From the absorption line measurement in LLIV1 against 4C~+66.09, we find spin temperature $\Trs \sim 75$\,K, cold gas mass fraction $f\sim0.5$ (with no component associated with a thermally unstable medium), and turbulent sonic Mach number $M_t\sim3.4$, characteristic of supersonic turbulence in the cold phase.
    \item
    Similar to the absorption line modeling against 4C~+66.09, our best emission line decomposition model has no unstable gas across the whole field of view, suggesting that the thermal condensation and phase transition are not on-going but rather have reached an equilibrium state.
    \item The cold phase of LLIV1 appears as a collection of elongated filaments that forms a closed structure within the field decomposed. 
    These substructures follow the orientation of the overall large scale cloud, along the diagonal of the GHIGLS field from north-west to south-east (in Galactic coordinates).
    \item The column density of the more diffuse warm phase is highest within the contour delimiting the presence of cold gas, but also exists outside of LLIV1.
    \item The angular power spectrum of the cold phase is slightly shallower than that of the warm phase, quantifying that the cold phases have relatively more structure on small scales.
    \item Our spatially resolved map of the cold gas mass fraction in LLIV1 is consistent with the absorption measurement against 4C~+66.09 and reveals significant variations 
    spanning the possible range of $f$, with mean and standard deviation of 0.33 and 0.19, respectively.
\end{itemize}

\acknowledgments
We  acknowledge  support  from  the  Natural  Sciences and Engineering Research Council (NSERC) of Canada. 
Some of this work was carried out when LV and MT participated in the Summer Undergraduate Research Program (SURP) hosted by the University of Toronto.
This research made use of the NASA Astrophysics Data System.

\vspace{5mm}

\software{Matplotlib \citep{hunter_2007}, NumPy \citep{van_der_walt_2011}, and AstroPy\footnote{\url{http://www.astropy.org}}, a community-developed core Python package for Astronomy \citep{astropy_2013, astropy_2018}, SciPy \citep{SciPy-NMeth}, scikit-image \citep{scikit-image}.}

\clearpage

\appendix

\restartappendixnumbering

\section{Parameter maps of individual Gaussian components (GHIGLS)}
\label{app:maps}

\begin{figure}
    \centering
    \includegraphics[width=\linewidth]{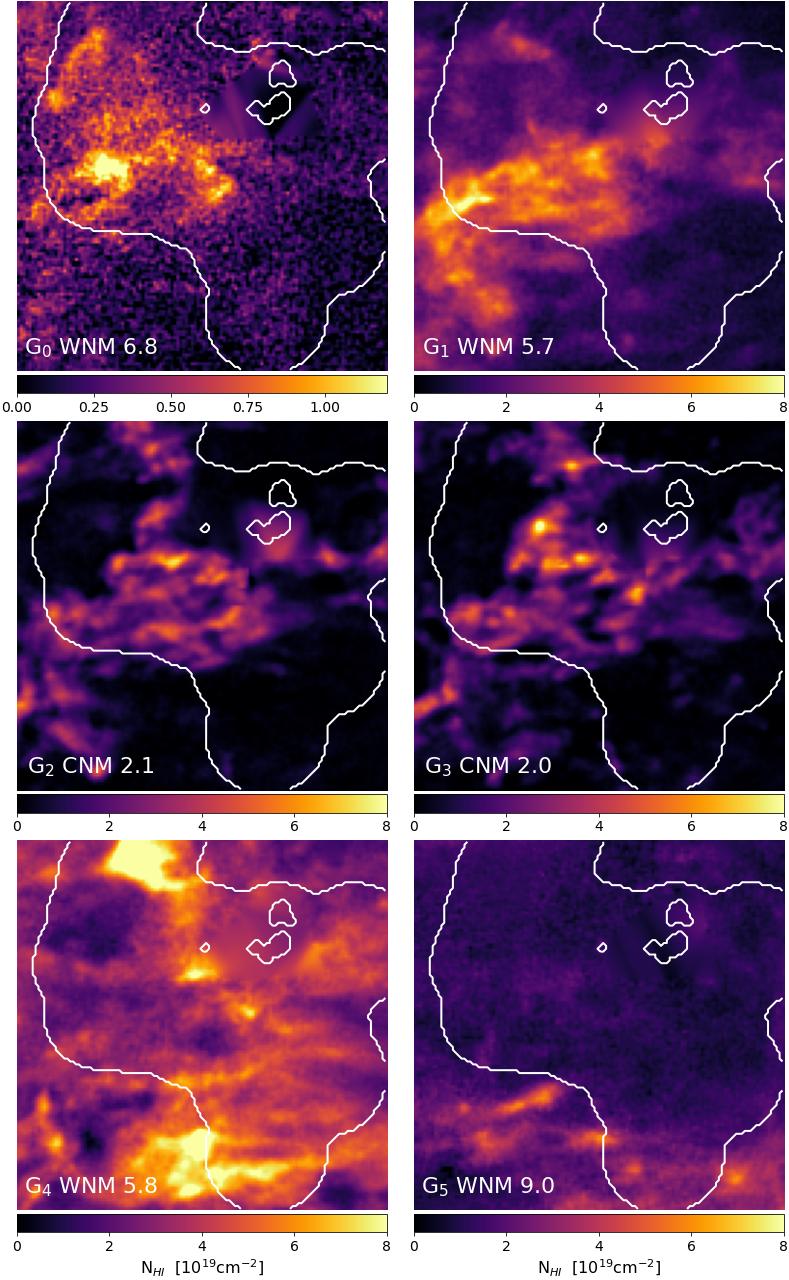}
    \caption{Mosaic of \nh\ maps of the individual Gaussian components. The maps are sorted by mean velocity. The white contour outlines the DHIGLS UM field 
    (Section \ref{subsec:data-dhigls}). The labels in the bottom left corner represent the classification of that particular Gaussian component. The number is the mean velocity dispersion ($\sigma$) in units of \kms.  See Table \ref{table:mean_var} and Figure \ref{fig:2d_hist}, which shows that $G_1$ -- $G_4$ are the components of interest for the IVC gas in LLIV1. The range of the color bar is the same for $G_1$ -- $G_5$, chosen to reveal the spatial structure and to emphasize the relative importance of the components across the region. Component $G_0$ is very weak and the color bar chosen has a much smaller range.
    }
    \label{fig:col-dens-mosaic}
\end{figure}

\begin{figure}
    \centering
    \includegraphics[width=\linewidth]{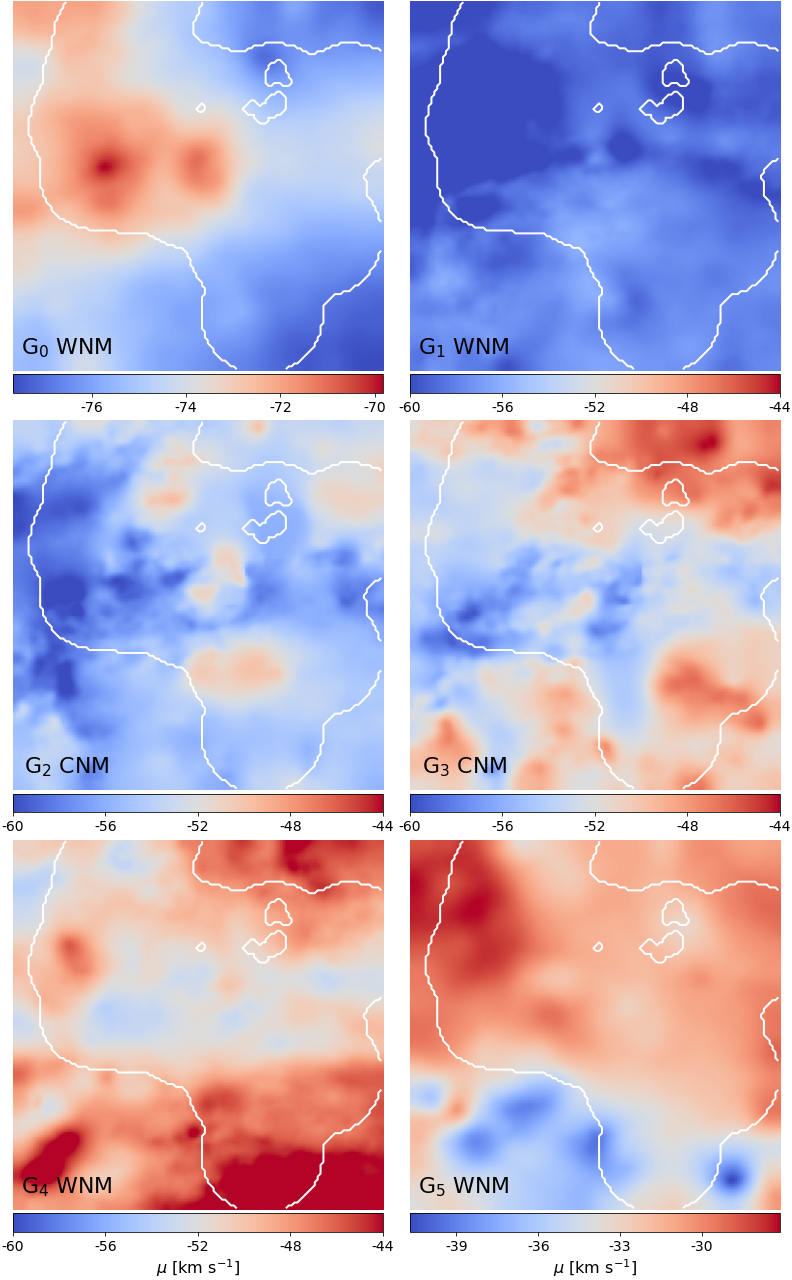}
    \caption{Mosaic like Fig.~\ref{fig:col-dens-mosaic}, but of maps of the mean velocity $\mu$ of the Gaussian components. The range of the color bar is the same for $G_1$ -- $G_4$. However, $G_0$ and $G_5$ are disjoint and the ranges of their color bars are chosen to show the spatial structure.
    }
    \label{fig:vel-mosaic}
\end{figure}

\begin{figure}
    \centering
    \includegraphics[width=\linewidth]{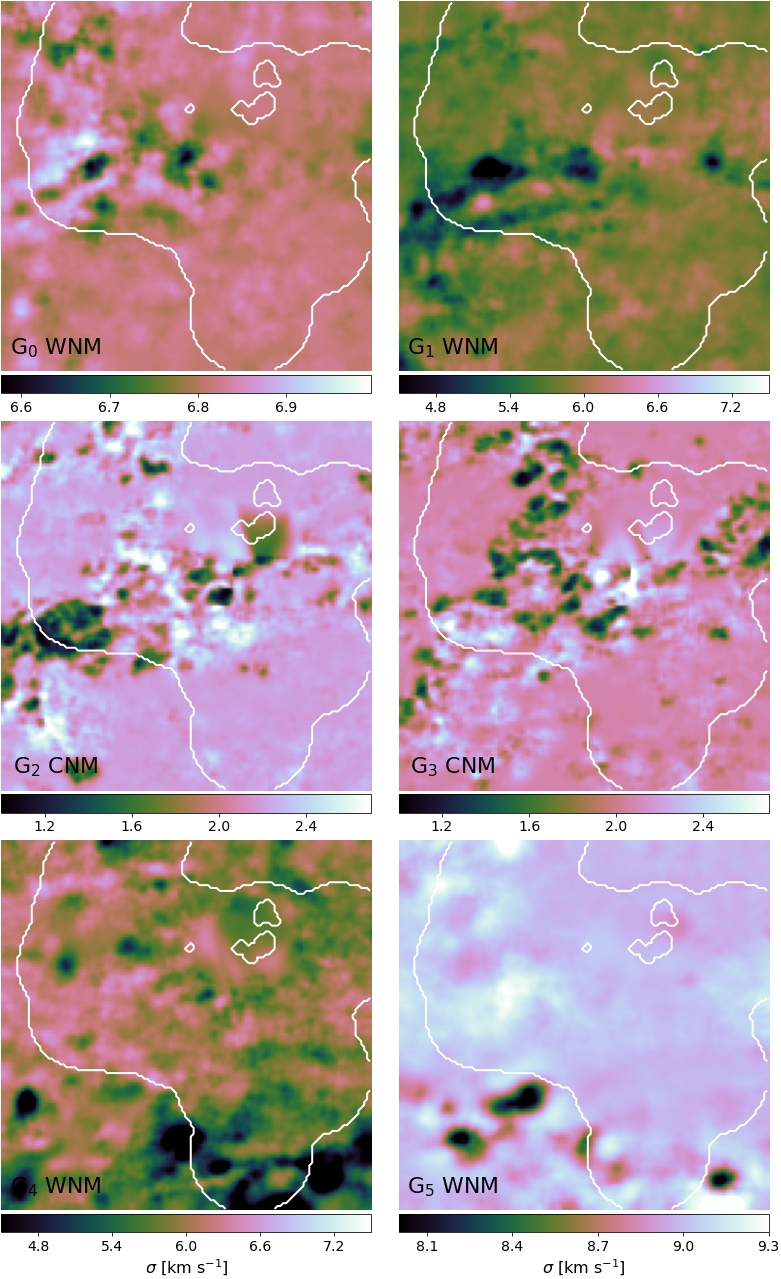}
    \caption{Mosaic like Fig.~\ref{fig:col-dens-mosaic}, but of maps of the velocity dispersion $\sigma$ of the Gaussian components.
    The range of the color bar is the same for the CNM components $G_2$ and $G_3$. Likewise, the WNM components $G_1$ and $G_4$ share the same range.  However, $G_0$ and $G_5$ are disjoint and the ranges of their color bars are chosen to show the spatial structure.
    }
    \label{fig:sigma-mosaic}
\end{figure}

Figure~\ref{fig:col-dens-mosaic} shows the column density maps of each Gaussian sorted by increasing column density-weighted mean velocities (see Table~\ref{table:mean_var}). Figures~\ref{fig:vel-mosaic} and \ref{fig:sigma-mosaic} shows the corresponding velocity fields and velocity dispersion fields. Note the shifting ranges of the color bars, as described in the captions.

The labels overlaid on the panels in Figure~\ref{fig:col-dens-mosaic} associate them with the clusters in Figure~\ref{fig:2d_hist}; e.g., for $G_2$ (in row 2, column 1), ``$G_2$ CNM 2.1'' indicates the number of the Gaussian component, the thermal phase, and the column density-weighted mean velocity dispersion $\left<\sigmab_n\right>$ in\,\kms\ as tabulated in Table~\ref{table:mean_var}.  $G_1$ -- $G_4$ are the components of interest for the IVC gas in LLIV1. $G_0$ and $G_5$ are important for the \ROHSA\ decomposition, but not particularly meaningful for the IVC analysis.

The white contour outlines the DHIGLS UM field 
    (Section \ref{subsec:data-dhigls} and Appendix \ref{app:mapsdh}) and also serves a fiducial role spatially.

\section{Impact of spatial resolution on the line}
\label{app:conv}
\setcounter{figure}{0}

\begin{figure}
    \begin{center}
        \includegraphics[width=\linewidth]{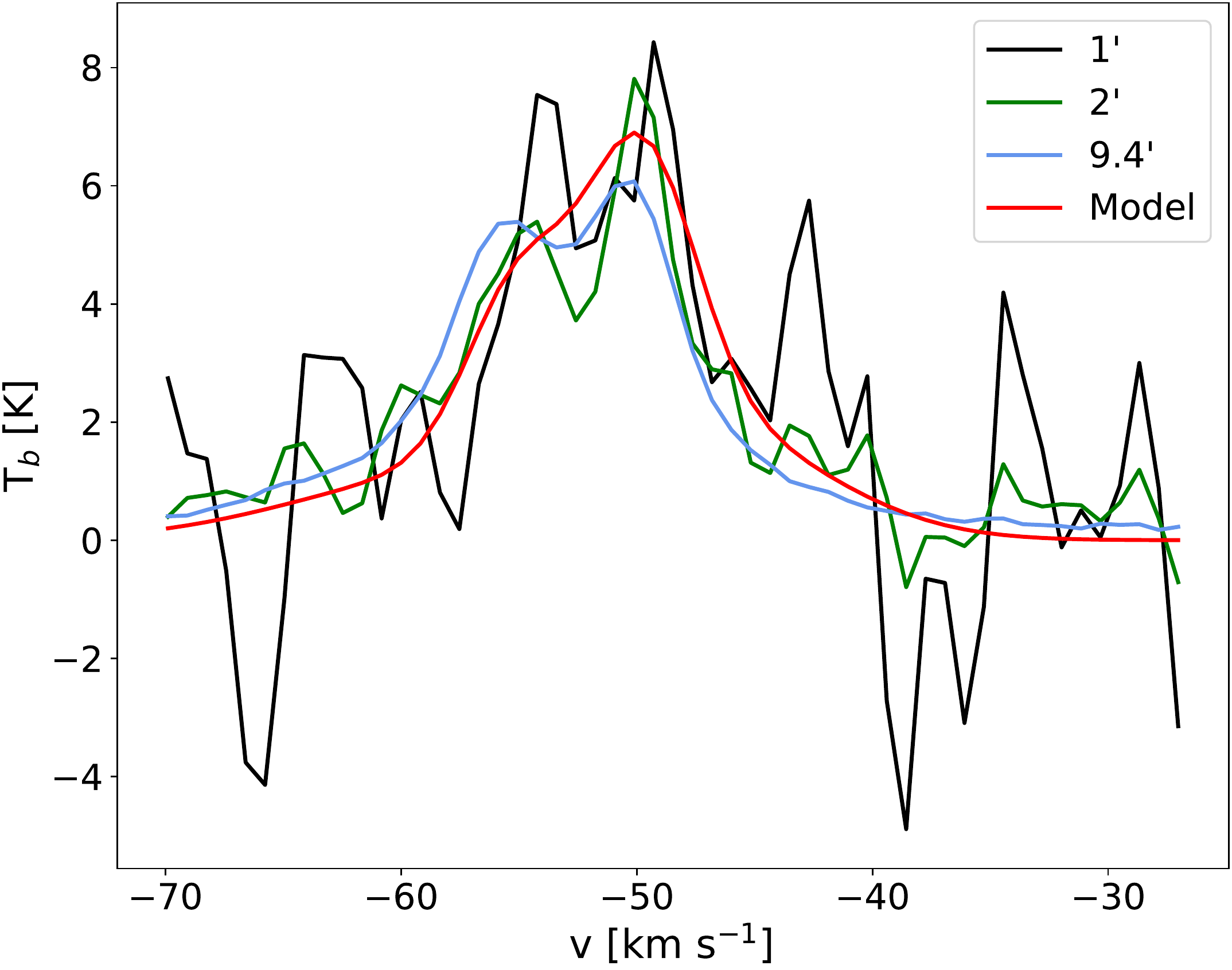}
        \caption{Spectrum showing a double peak of cold gas within the DHIGLS field. The black curve shows the original DHIGLS data, and the green and blue curves show the same data  convolved to $2^\prime$ and 9\farcm4, respectively. The red curve shows the model fit to the original DHIGLS data at $1^\prime$ resolution.}
        \label{fig:multirescon}
    \end{center}
\end{figure}

Figure~\ref{fig:multirescon} illustrates the impact of spatial resolution on a line of sight within the DHIGLS field chosen for its distinct double peak property. The black curve shows the original DHIGLS data, and the green and blue curves show the same data convolved to $2^\prime$ and 9\farcm4, respectively.
The \ROHSA\ model fit to the original DHIGLS data at $1^\prime$ resolution (red curve) captures the double-peak structure using two narrow Gaussians.

\section{Parameter maps of individual Gaussian components (DHIGLS)}
\label{app:mapsdh}
\setcounter{figure}{0}

\begin{figure*}
    \centering
    \includegraphics[width=\linewidth]
       {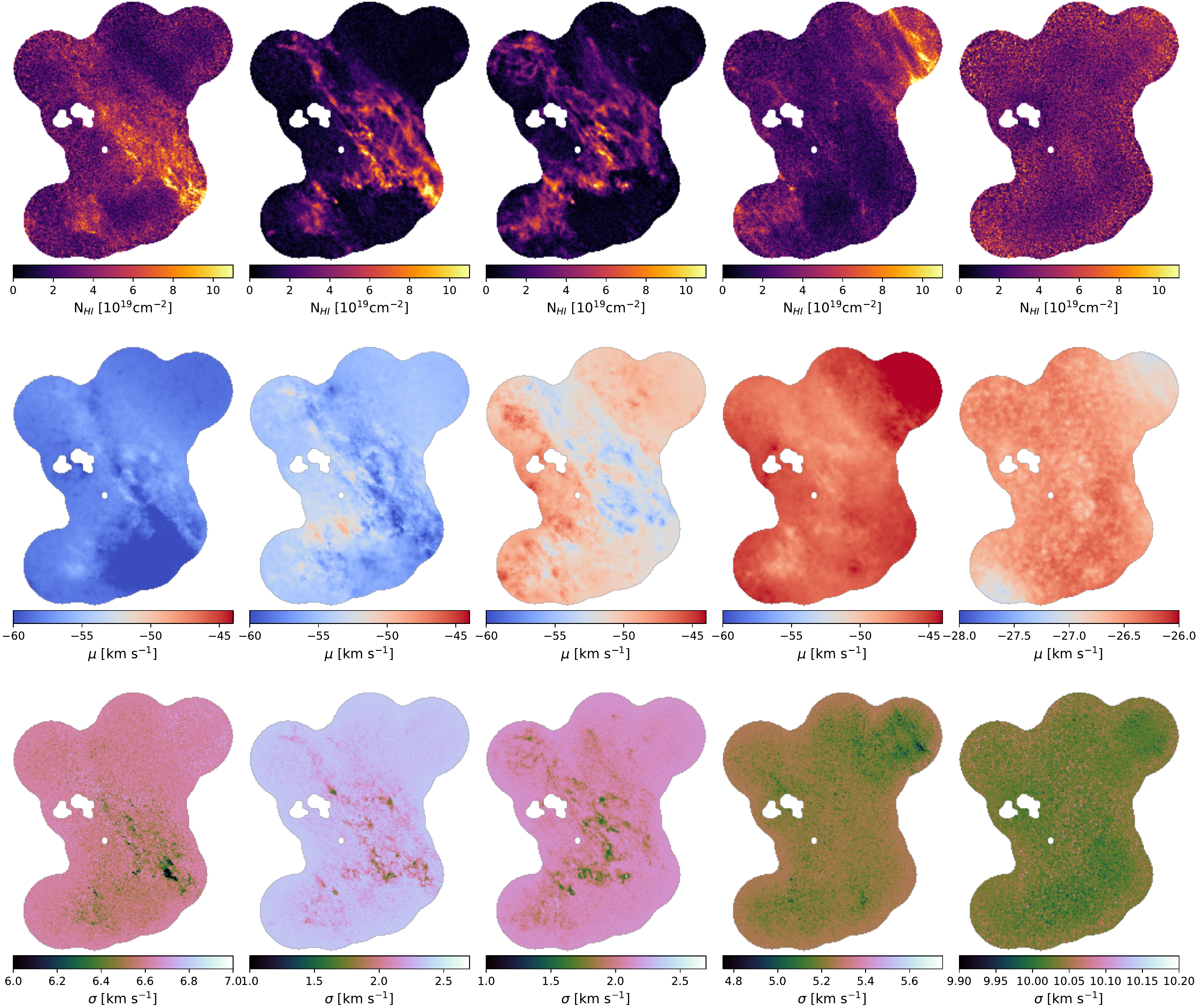}
    \caption{\nh\ and parameter maps of Gaussian components $G_1$ -- $G_5$ (Table \ref{table:mean_var}) identified in the analysis of IVC gas from the DHIGLS data, $G_1$ -- $G_4$ being the components of interest for LLIV1. 
    Top: column density \nh, all with the same color bar range. 
    Middle: corresponding mean velocity field $\mu$. The range of the color bar is the same for $G_1$ -- $G_4$ and identical to that in Figure \ref{fig:vel-mosaic}. $G_5$ is disjoint. 
    Bottom: velocity dispersion $\sigma$.  The range of the color bar is the same for the CNM components $G_2$ and $G_3$ and identical to that in Figure \ref{fig:sigma-mosaic}. The WNM components $G_1$ and $G_4$ are sufficiently different that we chose separate narrow ranges, both within the common range for the $G_1$ and $G_4$ components in Figure \ref{fig:sigma-mosaic}, so as not to suppress spatial structure.  $G_5$ is disjoint.
    }
    \label{fig:mosaic_params}
\end{figure*}

These are all arrayed in Figure \ref{fig:mosaic_params}. Note the shifting color bars. $G_1$ -- $G_4$ are the components of interest for the IVC gas in LLIV1. $G_5$ is important for the \ROHSA\ decomposition, but not particularly meaningful for the IVC analysis.

\clearpage
\bibliography{main}
\end{document}